\documentclass[12pt,draftclsnofoot,onecolumn]{IEEEtran}
 \usepackage[english]{babel}
 \usepackage{fancyhdr}
\usepackage{epsfig}
\usepackage{threeparttable}
\usepackage{epsf,epsfig}
\usepackage{amsthm}
\usepackage{amsmath}
\usepackage{amssymb}
\usepackage{amsfonts}
\usepackage[noadjust]{cite}
\usepackage{dsfont}
\usepackage{graphicx,amssymb,amsmath}
\usepackage{xcolor}
\usepackage{soul}
\usepackage{algorithmic,algorithm}
\usepackage{url}
\usepackage{subcaption}
\usepackage{bm}
\usepackage{mathtools}

\newtheorem{theorem}{Theorem}

\newtheorem{lemma}{Lemma}

\newtheorem{proposition}{Proposition}

\newtheorem{remark}{\bf Remark}

\def\proof{\noindent{\emph{Proof:} }}

\def\phi{\varphi}

\def\({\left(}
\def\){\right)}

\def\b0{{\mathbf{0}}}

\newcommand{\nn}{\nonumber}

\begin{document}
\graphicspath{{figure/}}
\title{\Large Joint  Parameter-and-Bandwidth Allocation for  Improving  the  Efficiency of  Partitioned Edge  Learning}
\author{Dingzhu Wen, Mehdi Bennis, and Kaibin Huang     \thanks{\setlength{\baselineskip}{13pt} \noindent D. Wen and K. Huang are with  The  University of  Hong Kong, Hong Kong.  M. Bennis is with University of Oulu, Finland. Corresponding email: huangkb@eee.hku.hk. }}
\maketitle
\vspace{-30pt}
\begin{abstract}
To leverage data and computation capabilities of mobile devices, machine learning algorithms are deployed at the network edge  for training \emph{artificial intelligence} (AI) models, resulting in the new paradigm of edge learning. In this paper, we consider the framework of \emph{partitioned edge learning} for iteratively training a large-scale model using many resource-constrained devices (called workers). To this end, in each iteration, the model is dynamically  partitioned into  parametric  blocks, which are downloaded to  worker groups for updating  using data subsets. Then, the local updates are uploaded to and cascaded by the server for updating a global model.  To reduce resource usage by minimizing the total learning-and-communication latency, this work focuses on the novel joint design of   parameter (computation load) allocation  and bandwidth allocation (for downloading and uploading). Two design approaches are adopted. First, a  practical sequential approach, called  partially integrated \emph{parameter-and-bandwidth allocation} (PABA),  yields two  schemes, namely \emph{bandwidth aware parameter allocation} and \emph{parameter  aware bandwidth allocation}. The former minimizes the load for the slowest (in computing) of worker groups, each training a  same  parametric block. The latter allocates the largest bandwidth to the worker being the latency bottleneck. Second, PABA are jointly optimized.  {Despite it being a \emph{nonconvex} problem,} an efficient and optimal solution algorithm  is derived by intelligently  nesting a bisection search and solving a \emph{convex} problem. Experimental results using real data demonstrate that integrating PABA  can substantially improve the performance of partitioned edge learning in terms of latency (by e.g., $46\%$) and accuracy (by e.g., $4\%$ {given the latency of 100 seconds}). 
\end{abstract}

\section{Introduction}
An enormous amount of data and computation resources are distributed over a large number of mobile devices \cite{gesbert2019guest}. This motivates the deployment of distributed machine learning algorithms at the network edge for fast and scalable model training. As a result, a new paradigm of computing, called edge machine learning, has emerged as an attractive and fast growing research {area} \cite{zhu2020toward,wang2018edge}. In this work, we consider  \emph{partitioned edge learning} (PARTEL) for large-scale model training. In the framework, a  model is partitioned into parametric  blocks that are downloaded onto devices  for distributed updating and the computation results are uploaded to a {server} for updating the global model \cite{li2013parameter}. This work aims at improving the efficiency of PARTEL by minimizing  the total communication-and-learning latency, thereby reducing the resource utilization. Under this objecive, a set of novel schemes are proposed by   jointly designing the model-partitioning based  load distribution over devices and  bandwidth allocation for their downloading and uploading.

\subsection{Partitioned Edge Learning}
\subsubsection{Partitioned  Learning}
A representative framework for partitioned distributed learning is called parameter server, which is  proposed in \cite{li2013parameter, li2014scaling} to distribute a large-scale learning task over many resource-constrained machines by \emph{model partitioning}. Implementing the classic method of \emph{block coordinate descent}  {(BCD)} \cite{wright2015coordinate},   the framework  iteratively and distributively  solves a large-scale model-optimization problem with a decomposable objective function [e.g., linear regression and \emph{support vector machine} (SVM)]. Specifically, a parameter {server} divides the model parameters into blocks and update each block in one iteration, called a \emph{communication round}. In each round, the server further divides and distributes the global dataset over workers so that they can locally compute the block gradients for updating the downloaded parametric block under their resource constraints. Then, the local gradients are uploaded to the server for aggregation (e.g., averaging) and updating the particular parametric block of the global model, completing the round. The framework is further developed in a series of work to reduce the learning latency by allowing overlapping of consecutive rounds  \cite{li2014communication} or workers to use a staled parametric block for computing their updates  \cite{ho2013more}.

Recent research on partitioned learning focus on \emph{convolutional neural network} (CNN) models. Such a model comprising nested layers does not have a decomposable (learning) loss function, making direct model partitioning sub-optimal.  Overcoming the limitation has driven researchers to extend the BCD method to CNN \cite{carreira2014distributed,zhang2016efficient}. Specifically, by introducing and conditioning on auxiliary variables, the layers in a CNN become conditionally independent and thus can be trained separately in different rounds. The cost due to a complex model architecture is that updating each layer (also a parametric block) requires multiple rounds instead of one as in the parameter-{server} framework  with a decomposable function. 

\subsubsection{Edge Implementation} While communication channels are abstracted as bit piles in prior work, PARTEL concerns the design of new communication techniques for efficient implementation of partitioned learning in wireless networks. This is an  uncharted area and the theme of this work. 
Connecting parameter servers with workers (edge devices) using wireless links gives rise to two challenges: 1) overcoming channel impairments (fading and noise) and scarcity of radio resources and 2) leveraging a massive number of resource-constrained devices for performing a single large-scale learning task. The direct application of traditional communication techniques may not be sufficient given the excessive communication overhead caused by large-scale model (with millions to billons of parameters) and large-scale dataset (typically comprising millions of high-dimensional multimedia samples). In this work, we adopt the new approach of integrated computation-and-communication design. Specifically, the model-and-data partitioning, computation load allocation, and radio resource allocation are jointly designed  so as to reduce the learning-and-communication latency, thereby minimizing the resource utilization. 

\subsection{ Federated  Edge Learning}
\subsubsection{Federated  Learning} 
Another mainstream framework for distributed learning, called \emph{federated learning}, was developed for the purpose of leveraging local data generated at edge devices while preserving their data privacy by avoiding direct data uploading \cite{lim2019federated}. The framework similar to parameter server but simpler as it involves no model or dataset partitioning. Implementing \emph{stochastic gradient descent} (SGD), federated learning requires each device to download and locally update the \emph{whole} model (or compute the needed gradient) and all devices to upload the computed models/gradients to update a global model after model/gradient aggregation at an edge server; the procedure iterates till the model converges. Communication efficiency is a main research theme in the area as excessive communication overhead is incurred by the repeated uploading of high-dimensional local models/gradients by many devices over many rounds. One approach is to reduce the number of uploading devices by allowing infrequent uploading by devices slow in computation \cite{chen2016revisiting}, or selecting those whose results are relatively more important for learning (see e.g. \cite{kamp2018efficient,chen2018lag}). 
Another approach is to directly compress local gradients exploiting their sparsity \cite{aji2017sparse}. 

\subsubsection{Edge Implementation}
Driven by the vision of edge intelligence, the new area of \emph{federated edge learning} (FEEL) has emerged, focusing on efficient implementations of federated learning in wireless networks. Based on the approach of communication-and-learning integration, many techniques are designed for efficient transportation of high-dimensional data over wireless channels. New multi-access schemes for FEEL, called ``over-the-air computing", are proposed in \cite{zhu2019broadband,amiri2019machine,ShiyuanmingAirComp} to support fast ``over-the-air" model/gradient aggregation using the waveform superposition property of a multi-access channel.  Another vein of research addresses the issue of \emph{radio resource management} (RRM) in FEEL systems such as bandwidth allocation \cite{chen2019joint}, multiuser scheduling \cite{yang2019scheduling}, and their joint design \cite{zeng2019energy,shi2019device}. Joint RRM and training batch-size selection is further investigated in \cite{ren2019accelerating} to accelerate the learning speed in FEEL systems. From the perspective of FEEL system performance, there exists a fundamental tradeoff between device energy consumption and learning speed, which is quantified in \cite{yang2019energy}. In addition, in view of the varying communication-and-computation capacities of different nodes, researchers have also developed a hierarchical network architecture for implementing large-scale FEEL \cite{abad2019hierarchical}.

\subsubsection{Federated vs. Partitioned Edge Learning}
The main  objective of FEEL is to exploit users' data without violating their privacy. The framework does not involve model partitioning and requires each  edge device to update a whole model.  Since the devices are resource constrained, FEEL is suitable for \emph{small-to-medium} learning tasks. In contrast, the objective of PARTEL  is to train a \emph{large-scale} model using many edge devices as workers via model partitioning. Therefore,  the design of efficient  PAETEL requires the integration of the partitioning for load allocation with radio resource allocation, which is a new challenge not faced in the area of FEEL. 

\subsection{Contributions and Organization}
In this paper, we consider a single-cell wireless  system supporting PARTEL. In the system, the workers are grouped and each group is responsible for updating an assigned parametric block. Within one group, the global dataset is distributed over workers so that each resource-constrained worker need compute the block update using only a data subset. In each communication round, an edge server coordinates the learning process by performing the following operations: 
\begin{enumerate}
\item \emph{Parameter allocation}, referring to partitioning the model into parametric blocks for load allocation; 
\item \emph{Bandwidth allocation}, namely partitioning the bandwidth for downloading latest values of parameters to workers and uploading their updates on parametric blocks;
\item Cascading the uploaded block updates to update the global model. 
\end{enumerate} 
Such coordinated distributed learning introduces the constraint of \emph{synchronized updates} by devices. The rounds  are repeated till the model converges.

One way {for} improving the efficiency of PARTEL is to minimize the total (communication plus computation) latency so as to minimize the utilization of radio and device-computation resources. Under the constraint of synchronized updates, the total latency depends on both the communication and computation latency of all devices, which can be controlled by bandwidth and parameter allocation, respectively. This motivates the current work to make the first attempt on jointly designing  \emph{parameter-and-bandwidth allocation} (PABA) for PARTEL systems. Our approach is to formulate and solve latency minimization problems for optimizing the PABA policy for given heterogeneous channel states and device-computation capacities. The specific  contributions and findings are summarized as follows. 

First,  practical PABA schemes are proposed for the scenario of large-scale network with fast varying channels. In this scenario, the direct optimization of joint PABA is a challenging problem, which  is non-convex with many variables and requiring an iterative solution method (see the second contribution). Moreover,  the task of solving the problem has to be repeated whenever the channels change. {To overcome the difficulty, we propose the practical  \emph{partially integrated PABA} where the designs of the two functional blocks are sequential: the first block (either parameter or bandwidth allocation) is designed independently of the other and the second block is designed conditioned on the first.}
 This results in two simple PABA schemes,  summarized as follows. 

\begin{figure}[t]
\begin{minipage}{\textwidth}
\center
\includegraphics[width=1\textwidth]{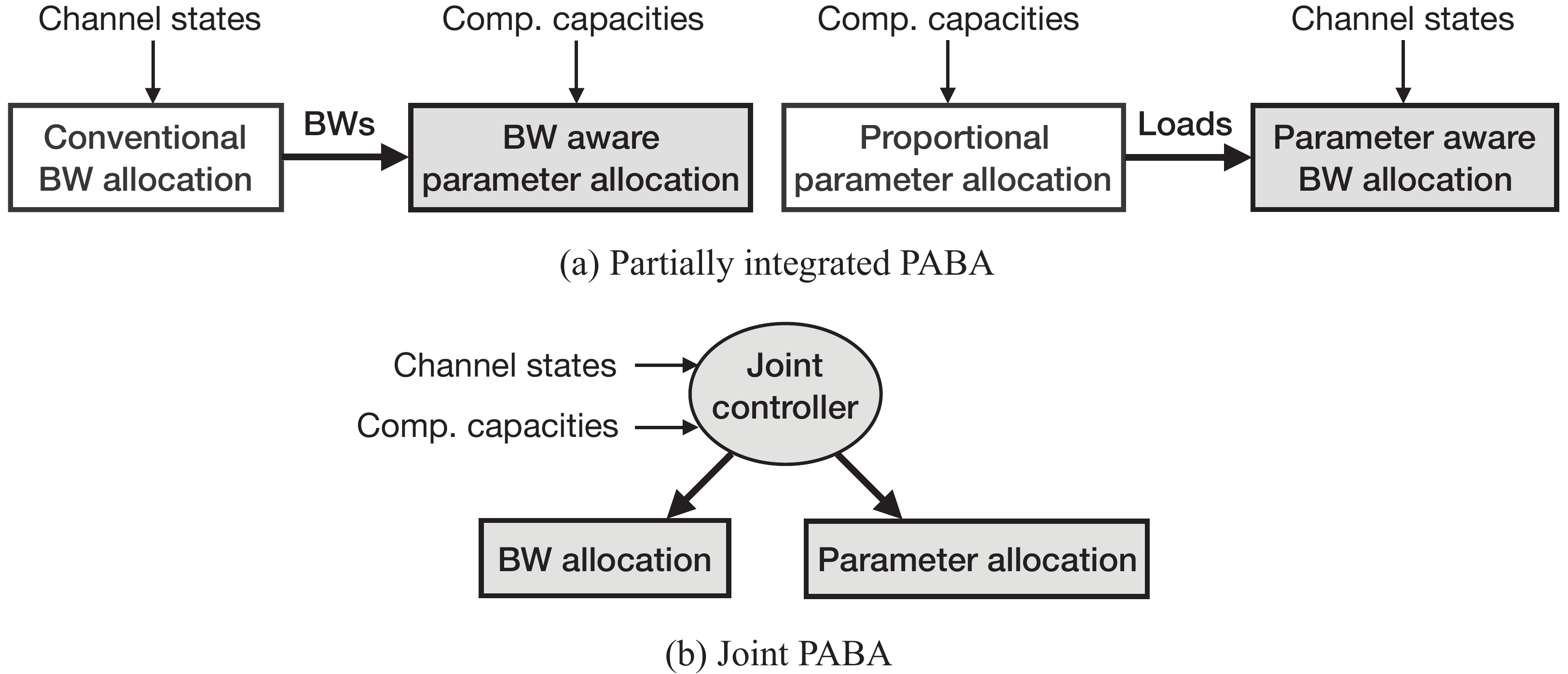}
\caption{Two approaches of designing parameter-allocation and bandwidth-allocation  blocks: (a) partially integrated design and (b) joint design.}\label{fig:JointDesigns}
\end{minipage}
\end{figure}

\begin{itemize}
\item {\bf Bandwidth aware parameter allocation}: Consider the optimization of the parameter allocation  conditioned on bandwidth allocation using a conventional scheme [see the left of Fig. \ref{fig:JointDesigns}(a)]. The optimization is shown be a linear program, allowing the optimal policy to be derived in closed form. The policy is  found to minimize the load for the worker group slowest in computation. To be precise, the parametric-block length assigned to one group is inversely proportional to its slowest worker's total latency. 

\item {\bf Parameter  aware bandwidth allocation}: Next, reversing the design order [see the right of Fig. \ref{fig:JointDesigns}(a)] yields the current scheme. By analyzing and exploiting the problem structure, solving the latency optimization problem is reduced to a simple bisection search. The resultant optimal policy for parameter  aware bandwidth allocation is  found  to allocate the largest bandwidth to the worker being the latency bottleneck to alleviate the bottleneck.
\end{itemize}

Next, targeting slowly varying channels, we develop an efficient iterative  algorithm for solving the problem of  joint PABA optimization, called \emph{fully integrated PABA}, as illustrated in Fig.~\ref{fig:JointDesigns}(b). To this end,  a useful  property is derived  that the optimal policy equalizes the \emph{group bandwidth allocation rates}, defined as the additional bandwidth required by assigning one additional parameter to the group for updating. Leveraging the property, an efficient solution method  is derived that intelligently nests a bisection search and solving a \emph{convex} problem. To gain further insights, two special cases with single-worker groups or  intra-group uniform computation capacities are considered. The optimal polices are derived in simple form and aligned with intuition (e.g., allowing more load to a group with better computation capacity).

The remainder of the  paper is organized as follows. In Section II, the system model is  introduced. In Section III, the total-latency minimization problem is formulated and  simplified. In Section IV and V, two schemes of partially integrated PABA  are designed while the scheme of fully integrated PABA is derived  in Section VI. Section VII presents the experimental results followed by concluding remarks in Section VIII.

\section{Models and Metrics}

\subsection{System  Model}
\begin{figure}[t]
\centering
\includegraphics[width=\textwidth]{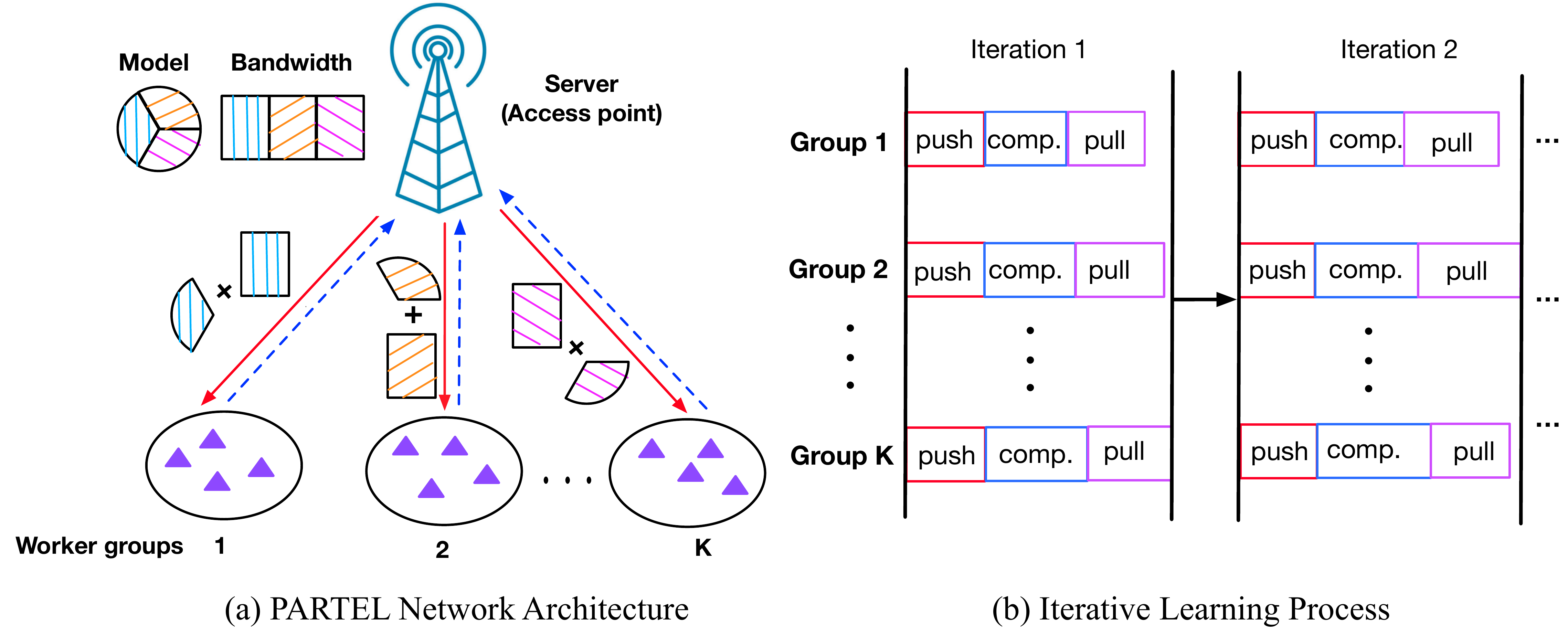}
\caption{System model and operations of the PARTEL framework.}\label{fig:SystemModel}
\end{figure}

A single-cell system is considered, as illustrated in Fig. \ref{fig:SystemModel}(a). In the cell, there are a server equipped with a single-antenna \emph{access point} (AP) and multiple single-antenna edge devices, serving as workers. The workers are divided into $K$ groups, identified by the index set $\mathcal{G}_1$, $\mathcal{G}_2$, ..., $\mathcal{G}_K$, each of which collaboratively performs one task. The $n$-th worker in group $\mathcal{G}_k$ is denoted as $(k,n)$. The server is connected to workers via wireless links. For simplicity, {the channels are assumed to be static in one iteration of model training and vary over different iterations.} We assume that the AP has the \emph{channel state information} (CSI) of all links that are useful for bandwidth allocation. The uplink/downlink spectrum is divided into orthogonal frequency non-selective channels, each of which is assigned to one worker. The downlink and uplink channel gains of worker $(k,n)$ are denoted as $H_{{\rm d},k,n}$ and $H_{{\rm u},k,n}$, respectively. 


\subsection{Learning Model}\label{sect:WDLA}
The PARTEL framework is designed for a large-scale learning task with a decomposable objective function. As mentioned, this is natural for algorithms such as  SVM and logistic regression \cite{wright2015coordinate} and can be made feasible for  CNN models using the  method of auxiliary variables \cite{carreira2014distributed,zhang2016efficient}. Following the literature, a decomposable  objective function has the following form (see e.g., \cite{wright2015coordinate}):
\begin{equation}\label{eq:ObjLoss}
\mathcal{L}({\bm \theta})  = \mathcal{F}({\bm \theta}) + \mathcal{R}({\bm \theta}),
\end{equation}
where ${\bm \theta}=\{\theta_1,...,\theta_{n_{\rm p}},...\theta_{N_{\rm p}}\}^T$ is the parameter vector of the learning model, $\mathcal{F}({\bm \theta})$ is the loss function, and $\mathcal{R}({\bm \theta})$ is the block-separable regularized function (e.g., $\ell_1$ and $\ell_2$ regularizations) to reduce the overfitting or increase the sparsity of the trained learning model, respectively. Specifically, the loss function can be written as
$\mathcal{F}({\bm \theta}) = \dfrac{1}{M} \sum\limits_{m=1}^M\phi ({\bm \theta};{\bf x}_m)$, where $\mathcal{X}$ = $\{{\bf x}_m\}$ is the dataset, $M$ is the size of the dataset, and $\phi(\cdot)$ is a smooth function. And the block-separable regularized function can be written as
$\mathcal{R}({\bm \theta}) = \sum\limits_{n_{\rm p}=1}^{N_{\rm p}} \psi(\theta_{n_{\rm p}})$,
where $\theta_{n_{\rm p}}$ is the $n_{\rm p}$-th element of ${\bm \theta}$ and $N_{\rm p}$ is the total number of parameters. { During training, if the regularization function $\mathcal{R}(\cdot)$ is smooth, gradient descent can be used for updating the learning model \cite{ruder2016overview}. Otherwise, the learning model is updated by another method, called proximal gradient descent \cite{parikh2014proximal}. }

\subsection{PARTEL Architecture}\label{sect:WDLA}
Consider the network architecture in  Fig. \ref{fig:SystemModel}(a). The global dataset is partitioned at the server and downloaded by workers such that each worker loads a data subset and each worker group has  the whole dataset. The model-parameter vector is partitioned into $K$ disjoint parametric blocks, as ${\bm \theta} = \{{\bm \theta}_1,...,{\bm \theta}_k,...,{\bm \theta}_K\}$, where ${\bm \theta}_k$ is assigned to group $\mathcal{G}_k$ for updating. One main benefit of PARTEL is that each resource-constrained worker {only needs to calculate} and transmit the gradient or proximal gradient of a parametric block over a data subset instead of the whole parameter vector during each iteration. As an example, considering  the case of smooth regularization function, only the following block of gradient elements is required for computation and transmission by worker $(k,n)$, 
\begin{equation}\label{eq:BlockGradient}
\dfrac{\partial \mathcal{L}(\bm \theta)}{\partial \theta_{n_{\rm p}}} =\dfrac{1}{M} \sum\limits_{{\bf x}_m\in \mathcal{X}_{k,n} } \dfrac{\partial\phi ({\bm \theta};{\bf x}_m)}{\partial \theta_{n_{\rm p}} } + \dfrac{1}{|\mathcal{G}_k|}\dfrac{{\rm d} \psi( \theta_{n_{\rm p}})}{{\rm d}  \theta_{n_{\rm p}}},\;\forall \theta_{n_{\rm p}} \in {\bm \theta}_k, 
\end{equation}
where $\mathcal{X}_{k,n}$ is the data subset at  worker $(k,n)$, $|\mathcal{G}_k|$ represents the number of workers in group $\mathcal{G}_k$,  $\dfrac{\psi( \theta_{n_{\rm p}})}{|\mathcal{G}_k|}$ represents the regularization function for each worker, and ${\bm \theta}_k$ with the size of $b_k$ is the parametric block assigned to group $\mathcal{G}_k$, respectively. 


In the PARTEL framework, one training iteration {of the learning model} is called \emph{one (communication) round}. In each round, the server first shares the whole model-parameter vector ${\bm \theta}$ to all workers. Then, the gradient or proximal gradient of each parametric block is calculated by one worker group. Finally, the gradients or proximal gradients of all groups are uploaded to the server to update all parameters. Wireless links are used for sharing the whole model parameters and uploading the gradients. Thereby,  to train the distributed learning algorithms in PARTEL framework, there are three steps in one  round, as follows.
\begin{itemize}
\item \emph{Push}: The server (AP) broadcasts the whole model parameters ${\bm \theta}$ to all workers.
\item \emph{Computation}: Each worker computes the gradient or the proximal gradient of the its assigned parametric block based on its loaded data subset.
\item \emph{Pull}: All workers upload the gradients or proximal gradients of their corresponding parametric blocks to the server. The server aggregates the gradients or proximal gradients from all groups and updates the corresponding parametric block. 
\end{itemize}
The iterative process of the distributed learning algorithms is shown in Fig. \ref{fig:SystemModel}(b). From the figure, synchronized updates are required in each round. {Synchronized updates arise from the operation of gradient aggregation in the pull step and refer to the requirement that all local updates need to be received by the server before the global model can be updated. Consequently, all devices are allowed the same duration (per-round latency) for uploading their local gradients and thus synchronized in update transmission.}
Hence, the latency in the round is decided by the ``slowest" worker.  In the sequel, the latency of any one round is defined.

\begin{remark}[Learning Convergence Speed]
\emph{With model updating per round, the distributed learning using the PARTEL architecture is optimal in the sense of achieving the same learning performance as the centralized learning within a same  number of rounds (see Lemma \ref{lma:NIter}). 
However, the implementation of PARTEL in wireless network  makes it necessary to measure the learning duration/latency in second. The reason is that finite radio bandwidth  and resource-constrained workers cause significant  latency of communication and computation in  each round. Therefore, optimizing parameter allocation and bandwidth allocation is important for ensuring fast model convergence as in achieving targeted learning performance within a given duration measured in second. 
}
\end{remark}
\begin{remark}[Relation with FEEL]\label{Rmk:RelationFEEL}
\emph{In the case of only one worker group and hence no model partitioning, the PARTEL architecture reduces to that of FEEL (with uploading per round), {as all workers calculate the gradient of the whole parameter vector over a data subset.} }
\end{remark}

\subsection{Latency Model}

Consider an arbitrary communication round, say the $r$-th round, and an arbitrary worker, say worker $(k,n)$. The latency, denoted as $t_{k,n}^{(r)}$, is composed of three parts:
\begin{equation}\label{eq:WorkerL}
t_{k,n}^{(r)} = T_{\rm ph}^{(r)} + \hat{t}_{k,n}^{(r)} + t_{{\rm pl},k,n}^{(r)},
\end{equation}
where $T_{\rm ph}^{(r)}$,  $\hat{t}_{k,n}^{(r)}$, and $t_{{\rm pl},k,n}^{(r)}$ correspond to the three steps, namely push, computation, and pull, respectively. 

\subsubsection{Push latency} The push latency is defined as the time for the server to broadcast the whole parameter vector ${\bm \theta}$ to all workers, which  is given by
\begin{equation}\label{eq:PushLa}
T_{\rm ph}^{(r)} = \max\limits_{\{(k,n)\}} \dfrac{A_{\rm p} N_{\rm p}}{BR_{{\rm d},k,n}^{(r)}},
\end{equation}
where $A_{\rm p}$ is the number of bits per model parameter, $N_{\rm p}$ is the total number of parameters, $B$ is the system bandwidth, and $R_{{\rm d},k,n}^{(r)}$ is the downlink spectrum efficiency of worker $(k,n)$. The efficiency can be written as $R_{{\rm d},k,n}^{(r)} = \log_2\left(1+P_{\rm b}H_{{\rm d},k,n}^{(r)}/N_0\right)$, 
where $P_{\rm b}$ is the transmission power of the AP, $H_{{\rm d},k,n}^{(r)}$ is the downlink channel gain, and $N_0$ is the channel noise variance. Note that the push latency is a constant identical for all workers.

\subsubsection{Computation latency} Denote the number of computation operations to calculate the gradient with respect to one parameter using one data sample as $O$, the number of data samples loaded by worker $(k,n)$ as $D_{k,n}$, and the CPU frequency  of worker $(k,n)$ as $f_{k,n}^{\rm c}$. Then,  the computation latency of worker $(k,n)$ is a function of its assigned load $b_k^{(r)}$, i.e., the length of its assigned parametric block ${\bm \theta}_k^{(r)}$. It can be written as
\begin{equation}\label{eq:CompL}
\hat{t}_{k,n}^{(r)}\left(b_k^{(r)}\right) = \dfrac{b_k^{(r)} D_{k,n} O}{f_{k,n}^{\rm c}}.
\end{equation}

\subsubsection{Pull latency} The pull latency consists of two parts. One is the time for worker $(k,n)$ to upload the gradients to the server, denoted as $\tilde{t}^{(r)}_{k,n}$. The other is the time for the server to update the learning model, denoted as $T_{\rm s}^{(r)}$. Hence, the pull latency is given by
\begin{equation}\label{eq:PullL}
t_{{\rm pl},k,n}^{(r)} = \tilde{t}^{(r)}_{k,n}+T_{\rm s}^{(r)}.
\end{equation}
The time $T_{\rm s}^{(r)}$ is the same for all workers. The uploading time $\tilde{t}^{(r)}_{k,n}$ is given by 
\begin{equation}
\tilde{t}_{k,n}^{(r)}\left(b_k^{(r)},\rho_{k,n}^{(r)}\right) = \dfrac{ A_{\rm g} b_k^{(r)}}{\rho_{k,n}^{(r)} B R_{{\rm u},k,n}^{(r)}},
\end{equation}
where $A_{\rm g}$ is the number of bits for each gradient element, $b_k^{(r)}$ is the assigned parametric-block length, $\rho_{k,n}^{(r)}$ is the ratio of uplink bandwidth allocated to worker $(k,n)$, and $R_{{\rm u},k,n}^{(r)}$ is the uplink spectrum efficiency. The efficiency is $R_{{\rm u},k,n}^{(r)} = \log_2\left(1+P_{\rm u}H_{{\rm u},k,n}^{(r)}/N_0\right)$, 
where $P_{\rm u}$ is the uplink transmission power and $H_{{\rm u},k,n}^{(r)}$ is the uplink channel gain.

We define the group latency in the $r$-th round as follows.  Since all parameters should be updated in the round, the group latency is decided by the ``slowest" worker. The latency of group $\mathcal{G}_k$ is thus given as:
\begin{equation}\label{eq:GroupL}
t_k^{(r)}\left(b_k^{(r)},\{\rho_{k,n}^{(r)}\}\right) = \max\limits_{n\in \mathcal{G}_k}\; t_{k,n}^{(r)}\left(b_k^{(r)},\rho_{k,n}^{(r)}\right),
\end{equation}
where $t_{k,n}^{(r)}\left(b_k^{(r)},\rho_{k,n}^{(r)}\right)$ is the latency of worker $(k,n)$ defined in \eqref{eq:WorkerL} and its three components are described in \eqref{eq:PushLa}, \eqref{eq:CompL}, and \eqref{eq:PullL}.

Next, we define the total latency in the $r$-th round. Given synchronized updates, the total latency in this communication round depends on the ``slowest" group:
\begin{equation}\label{eq:OverallL}
t^{(r)}\left(\{b_k^{(r)}\},\{\rho_{k,n}^{(r)}\}\right) = \max\limits_{k}\; t_k^{(r)}\left(b_k^{(r)},\{\rho_{k,n}^{(r)}\}\right).
\end{equation}
 
\section{Problem Formulation and Simplification}
Based on the models described in the preceding section, the problem of learning-latency minimization is formulated in this section under two constraints, one on the total number of parameters and the other on the total bandwidth. Then, the problem is simplified as an equivalent one-round problem.

\subsection{Problem Formulation}
The learning latency depends on two factors. One is the \emph{number of communication rounds required for model convergence}, denoted as $R$, and the other is the \emph{per-round latency}. Thus, the total learning latency is given as
\begin{equation}\label{eq:TrainTime}
t_{\rm learn} \left(\{b_k^{(r)}\},\{\rho_{k,n}^{(r)}\}\right) = \sum\limits_{r=1}^{R} t^{(r)} \left(\{b_k^{(r)}\},\{\rho_{k,n}^{(r)}\}\right),
\end{equation}
where $t^{(r)}\left(\{b_k^{(r)}\},\{\rho_{k,n}^{(r)}\}\right)$ is the latency of the $r$-th round defined in \eqref{eq:OverallL}. We aim at minimizing the learning latency by optimizing the distribution of parametric blocks, or called parameter allocation, and the bandwidth allocation. Parameter allocation must satisfy the following constraints on the total number of parameters: 
\begin{equation}\text{(C1.1: Parametric block constraints)}\;
\left\{
\begin{aligned}
&\sum\limits_{k=1}^K b_k^{(r)} = N_{\rm p},\;1\leq r\leq R,\\
&b_k^{(r)} \in \mathbb{Z}^{+},~\forall k,\;1\leq r\leq R,
\end{aligned}
\right.
\end{equation}
where $N_{\rm p}$ is the total number of parameters and $b_k^{(r)}$ is length of the parametric block assigned to group $\mathcal{G}_k$ in the $r$-th round. On the other hand, the bandwidth allocation should satisfy the following constraints on the total bandwidth:
\begin{equation}\text{(C1.2: Bandwidth constraints)}\;
\left\{
\begin{aligned}
&\sum\limits_{k=1}^K\sum\limits_{n\in \mathcal{G}_k} \rho_{k,n}^{(r)} \leq 1,\;1\leq r\leq R,\\
&\rho_{k,n}^{(r)} \geq 0 ,~\forall (k,n),\;1\leq r\leq R,
\end{aligned}
\right.
\end{equation}
where $\rho_{k,n}^{(r)}$ is the ratio of the bandwidth allocated to worker $(k,n)$ in the $r$-th round. Under the constraints, the problem of learning-latency minimization can be formulated as
\begin{equation}\text{({P1}: Learning-latency minimization)}\;
\begin{aligned}
\mathop{\min }\limits_{\{b_k^{(r)}\},\{\rho_{k,n}^{(r)}\}}\;  &\sum\limits_{r=1}^{R} t^{(r)} \left(\{b_k^{(r)}\},\{\rho_{k,n}^{(r)}\}\right) \\
{\text{s.t.}}\;\; &\text{(C1.1)} \; \& \; \text{(C1.2)}.
\end{aligned}
\end{equation}
In the sequel, we prove that ({P1}) can be reduced to an equivalent one-round problem. 

\subsection{Equivalent One-Round Latency Minimization}
As shown in the following lemma, the convergence rates (in rounds) of the learning algorithms implemented at PARTEL are equivalent to the corresponding centralized ones, where the whole training process, including gradient calculation and model updating, is performed at the server.
\begin{lemma}[How many communication rounds?]\label{lma:NIter}
\emph{
The distributed learning algorithms implemented at PARTEL are equivalent to the corresponding centralized ones in terms of convergence rate as measured by the required number of communication rounds (see e.g., \cite{parikh2014proximal}, for convergence analysis on the latter). Specifically, for distributed learning, the values of gradients or proximal gradients calculated in each round and the number of rounds required for model convergence are independent on parameter allocation and bandwidth allocation. 
}
\end{lemma}
\proof See Appendix \ref{apdx:NIter}.

The following proposition follows from Lemma \ref{lma:NIter}.
\begin{proposition}[Problem simplification]\label{thm:EqLatency}
\emph{
The learning-latency-minimization problem in ({P1}) is equivalent to separately minimizing the latencies for all rounds:
\begin{equation}\text{({P2}: One-round-latency minimization)}\;
\begin{aligned}
\mathop{\min }\limits_{\{b_k^{(r)}\},\{\rho_{k,n}^{(r)}\}}\;  &t^{(r)}\left(\{b_k^{(r)}\},\{\rho_{k,n}^{(r)}\}\right)  \\
{\text{s.t.}}\;\; & \text{(C1.1)} \; \& \; \text{(C1.2)}.
\end{aligned}
\end{equation}
}
\end{proposition}

The simplified problem is solved in the following sections to obtain the optimal polices for PABA. For simplicity, the notation $(r)$ is omitted.

\section{Bandwidth Aware Parameter Allocation}
In this section, the scheme of bandwidth aware parameter allocation is designed based on the approach on the left of Fig. \ref{fig:JointDesigns}(a). Given bandwidth allocation, the optimal parameter allocation is proposed, which requires the latencies of all groups equal to the optimum. Besides, according to the optimal solution, the length of the parametric block assigned to group $\mathcal{G}_k$ is inversely proportional to its slowest worker's total latency for computing and uploading one gradient element.

First, the bandwidths are allocated to the workers independent of their assigned parametric block, e.g., equal bandwidth allocation. Next, given allocated bandwidths, the parameters are allocated by solving  ({P2}), giving the algorithm of bandwidth aware parameter allocation. Specifically, given the bandwidth-allocation scheme $\{\rho_{k,n}^*\}$, the problem of one-round-latency minimization in ({P2}) can be simplified as
\begin{equation}\text{({P3}: Parameter allocation})\;
\begin{aligned}
\mathop{\min }\limits_{\{b_k\}}\mathop{\max }\limits_k\;  &t_k\left(b_k\right), \\
{\text{s.t.}}\;\; & b_k \in \mathbb{Z}^{+},\;1\leq k\leq K,\\
&\sum\limits_{k=1}^K b_k = N_{\rm p},
\end{aligned}
\end{equation}
where $t_k\left(b_k\right)$ and $b_k$ are the latency and the parametric-block length of group $\mathcal{G}_k$, respectively.
By substituting the push latency in \eqref{eq:PushLa}, the computation latency in \eqref{eq:CompL}, and the push latency in \eqref{eq:PullL} into the group latency in \eqref{eq:GroupL}, it can be derived that
\begin{equation}\label{eq:GroupLinear}
\begin{aligned}
t_k\left(b_k\right) = &\max\limits_{n\in \mathcal{G}_k} t_{k,n}= T_{\rm ph}+\max\limits_{n\in \mathcal{G}_k}\left\{\dfrac{D_{k,n} O}{f_{k,n}^{\rm c}} + \dfrac{ A_{\rm g} }{\rho_{k,n}^*BR_{{\rm u},k,n}}\right\}b_k+T_{\rm s},
\end{aligned}
\end{equation}
which shows that $t_k\left(b_k\right)$ is a linear function of $b_k$. 
Furthermore, by defining $t_{\rm PA}=\max\nolimits_{k} t_k\left(b_k\right)$, ({P3}) can be converted into the following \emph{mixed-integer linear problem} (MILP),
\begin{equation}\label{eq:MILP}
\begin{aligned}
\mathop{\min }\limits_{\{b_k\},t_{\rm PA}}\;  &t_{\rm PA}, \\
{\text{s.t.}}\;\; & b_k \in \mathbb{Z}^{+},\;1\leq k\leq K,\\
&\sum\limits_{k=1}^K b_k = N_{\rm p},\\
&t_k\left(\{b_k\}\right)\leq t_{\rm PA},\;1\leq k\leq K.
\end{aligned}
\end{equation}
To solve \eqref{eq:MILP}, we follow the typical way, which first relaxes $\{b_k\}$ to be continuous and then round the solution. The error caused by the relaxation and rounding is just one parameter and is negligible due to the typically large values of $\{b_k\}$ (e.g., thousands to tens of thousands). 
\begin{theorem}[Relaxed parameter allocation]\label{thm:RBS}
\emph{
By relaxing the integer constraints $\{b_k \in \mathbb{Z}^{+},\;1\leq k\leq K\}$ to $\{b_k\geq 0,\;1\leq k\leq K\}$, the problem in \eqref{eq:MILP} can be solved by linear programming. The solution requires all groups to have the same latency: 
\begin{equation}
t_k\left(b_k\right)= t_{\rm PA}^*,\; \forall k,
\end{equation}
where $t_{\rm PA}^*$ solves the following equation and can be computed using e.g., a bisection search:
\begin{equation}\label{eq:OPBSL}
\sum\limits_{k=1}^K \dfrac{ t_{\rm PA}^*-T_{\rm ph}-T_{\rm s} }{\max\limits_{n\in \mathcal{G}_k}\left\{\dfrac{D_{k,n} O}{f_{k,n}^{\rm c}} + \dfrac{ A_{\rm g}}{\rho_{k,n}^* B R_{{\rm u},k,n}}\right\}}=N_{\rm p}.
\end{equation}
The optimal parameter-allocation policy assigns $b_k^*$ parameters to group $\mathcal{G}_k$ with $b_k^*$ given by
\begin{equation}\label{eq:RBS}
b_k^* = \dfrac{ t_{\rm PA}^*-T_{\rm ph}-T_{\rm s} }{\max\limits_{n\in \mathcal{G}_k}\left\{\dfrac{D_{k,n} O}{f_{k,n}^{\rm c}} + \dfrac{ A_{\rm g} }{\rho_{k,n}^*BR_{{\rm u},k,n}}\right\}},\;1\leq k\leq K.
\end{equation}
}
\end{theorem}
\proof See Appendix \ref{apdx:RBS}.

Two observations can be made from Theorem \ref{thm:RBS}. First, according to \eqref{eq:OPBSL}, \emph{the minimal per-round latency, $t_{\rm PA}^*$, linearly increases as the total number of parameters, $N_{\rm p}$, grows}. Second, in \eqref{eq:RBS}, the terms $\dfrac{D_{k,n} O}{f_{k,n}^{\rm c}}$ and $\dfrac{ A_{\rm g}}{\rho_{k,n}^*BR_{{\rm u},k,n}}$ are the time for computing one gradient element and the time for uploading the element for worker $(k,n)$, respectively. From \eqref{eq:RBS}, we can observe that \emph{the optimal parametric-block length assigned to group $\mathcal{G}_k$, say  $b_k$, is inversely proportional to its slowest worker's total latency for computing and uploading one gradient element}.

Rounding the real-valued numbers of parameters assigned to the groups gives the algorithm of bandwidth aware parameter allocation in Algorithm \ref{Ag:LLB}.
\begin{algorithm}[h]
\caption{Bandwidth Aware Parameter Allocation}\label{Ag:LLB}
1: {\bf Input:}  $\{\rho_{k,n}^{*}\}$, $\{R_{{\rm u},k,n} \}$.
\begin{itemize}
\item $\{\rho_{k,n}^{*}\}$, pre-determined bandwidth-allocation scheme,
\item $\{R_{{\rm u},k,n} \}$, the uplink spectrum efficiencies of all workers.
\end{itemize}

2: Get the optimal latency $t_{\rm PA}^*$ of the relaxed problem by solving \eqref{eq:OPBSL} with bisection method.

3: Determine the practical block size $\{\hat{b}_k^*\}$ as 

\begin{equation}\label{eq:PBS}
\left\{
\begin{aligned}
&\hat{b}_k^* ={\rm round} \left(\dfrac{ t_{\rm PA}^*-T_{\rm ph}-T_{\rm s} }{\max\limits_{n\in \mathcal{G}_k}\left\{\dfrac{D_{k,n} O}{f_{k,n}^{\rm c}} + \dfrac{ A_{\rm g} }{\rho_{k,n}^*BR_{{\rm u},k,n}}\right\}}\right),\; \forall k\in[1,K-1],\\
&\hat{b}_K^* =N_{\rm p}-\sum\limits_{k=1}^{K-1}\hat{b}_k.
\end{aligned}
\right.
\end{equation}

4: Calculate the near-optimal latency $\hat{t}_{\rm LB}^*$ with $\{\hat{b}_k^*\}$.

5: {\bf Output:} $\{\hat{b}_k^*\}$ and $\hat{t}_{\rm LB}^*$.
\end{algorithm}
\setlength{\textfloatsep}{1em} 

\section{Computation Aware Bandwidth Allocation}
In this section, the scheme of parameter aware bandwidth allocation {is designed based on} the  approach on the right of Fig. \ref{fig:JointDesigns}(a). Given parameter allocation, the optimal bandwidth allocation is proposed, where all workers' latencies equal to the optimum and most bandwidth should be allocated to the worker with smallest communication rate and longest computation time. 

First, the parametric-block lengths are assigned to the groups independent of their spectrum efficiencies, e.g., the parametric-block length assigned to one group is proportional to its computation latency of computing one gradient element. Next, given assigned parametric blocks, the bandwidths are allocated by solving ({P2}), giving the algorithm of parameter aware bandwidth allocation. Specifically, given the parameter-allocation policy $\{\hat{b}_k^*\}$, the problem of one-round-latency minimization in ({P2}) reduces to
\begin{equation}\text{({P4}: Bandwidth allocation)}\;
\begin{aligned}
\mathop{\min }\limits_{\{\rho_{k,n}\}}\mathop{\max }\limits_k\;  &t_k\left(\{\rho_{k,n}\}\right), \\
{\text{s.t.}}\;\; &\rho_{k,n}\geq 0 ,~\forall (k,n),\\
&\sum\limits_{k=1}^K\sum\limits_{n\in \mathcal{G}_k} \rho_{k,n} \leq 1,
\end{aligned}
\end{equation}
where $t_k\left(\{\rho_{k,n}\}\right)$ is the latency of group $\mathcal{G}_k$ defined in \eqref{eq:GroupL} and $\rho_{k,n}$ is the uplink bandwidth ratio allocated to worker $(k,n)$. 

\begin{lemma}[Convexity of bandwidth allocation]\label{lma:CBA}
\emph{
({P4}) is a convex problem.
}
\end{lemma}
\proof See Appendix \ref{apdx:CBA}.

By solving the convex problem, the minimal latency and the optimal bandwidth-allocation scheme can be obtained, as shown in the following theorem.
\begin{theorem}[Parameter aware bandwidth allocation]\label{thm:BWS}
\emph{
The optimal solution of the bandwidth-allocation problem in ({P4}) requires all workers have the same latency: $ t_{k,n}\left(\rho_{k,n}\right) = t_{\rm BA}^*,\; \forall (k,n)$,
where $t_{\rm BA}^*$ is the minimal latency that solves the following equation and can be computed using e.g., a bisection search:
\begin{equation}\label{eq:OPBW}
\sum\limits_{k=1}^K\sum\limits_{n\in \mathcal{G}_k} \dfrac{ \hat{b}_k^* A_{\rm g} }{  \left(t_{\rm BA}^* - T_{\rm s} - T_{\rm ph} - \hat{T}_{k,n}\right)R_{{\rm u},k,n} }=B,
\end{equation}
where $\{\hat{T}_{k,n} = \hat{t}_{k,n}(\hat{b}_k^*)\}$ are the computation latency and are constants. The resultant scheme of computation aware bandwidth allocation is given as
\begin{equation}\label{eq:BWS}
\rho_{k,n}^* = \dfrac{\hat{b}_k^* A_{\rm g}}{  \left(t_{\rm BA}^* - T_{\rm s} - T_{\rm ph} - \hat{T}_{k,n}\right)R_{{\rm u},k,n} B},\; \forall (k,n).
\end{equation}
}
\end{theorem}

\proof See Appendix \ref{apdx:BWS}.

Two observations can be made from Theorem \ref{thm:BWS}. First, in \eqref{eq:OPBW}, the system bandwidth is a strict decreasing function of the optimum $t_{\rm BA}^*$. In turn, it's easy to show that \emph{the optimal latency $t_{\rm BA}^*$ strictly decreases as the system bandwidth $B$ increases}. Second, for the optimal bandwidth-allocation scheme in \eqref{eq:BWS}, the allocated bandwidth of any one worker is a decreasing function of its uplink data rate and is an increasing function of its computation time. In other words, \emph{the most bandwidth should be allocated to the worker with smallest uplink rate and longest computation time}. 

The optimal scheme of parameter-aware bandwidth allocation is summarized in Algorithm \ref{Ag:BWS}.
\begin{algorithm}
\caption{Parameter Aware Bandwidth Allocation}\label{Ag:BWS}

1: {\bf Input:} $\{\hat{b}_k^*\}$ and $\{R_{{\rm u},k,n} \}$.
\begin{itemize}
\item $\{\hat{b}_k^*\}$, the pre-determined parametric-block lengths,
\item $\{R_{{\rm u},k,n} \}$, the uplink spectrum efficiencies of all workers.
\end{itemize}
  
2: Calculate the optimal latency $t_{\rm BA}^*$ by solving \eqref{eq:OPBW} with bisection method.

3: Determine the bandwidth $\{\rho_{k,n}^*\}$ as \eqref{eq:BWS}.

4: {\bf Output:} $\{\rho_{k,n}^*\}$ and $t_{\rm BA}^* $.
\end{algorithm}

\section{Joint Parameter Allocation and  Bandwidth Allocation}
In this section, joint PABA based on the design approach in Fig. \ref{fig:JointDesigns}(b) is considered. Leveraging the results in preceding sections, the optimization problem ({P2}) is simplified. This allows an efficient solution method to be developed for computing the optimal policy for joint PABA. To gain further insights, two special cases are considered.

\subsection{Optimal Joint PABA}

The optimal joint PABA policy  for PARTEL is computed and analzyed by solving Problem ({P2}) following a series of steps as follows. 

\subsubsection{Problem Simlification}

First, we simplify ({P2}) by using the results in Theorem \ref{thm:BWS} and relaxing the parametric-block lengths $\{b_k\}$ to be continuous. According to 
Theorem \ref{thm:BWS}, to achieve the minimal latency, all workers should have the same one-round latency [defined in \eqref{eq:OverallL}], namely $t_{k,n}=t,\;\forall (k,n)$. In {Theorem \ref{thm:BWS}}, $\{b_k\}$ are given but they are  variables in the current case. Thus, the bandwidth-allocation policy  in \eqref{eq:BWS} should be rewritten as a function of $\{b_k\}$: 
\begin{equation}\label{eq:Rho}
\rho_{k,n}\left(b_k,t\right) = \dfrac{b_k A_{\rm g}}{  \left[t - T_{\rm s} - T_{\rm ph} - \hat{t}_{k,n}\left(b_k\right) \right]R_{{\rm u},k,n} B},\; \forall (k,n),
\end{equation}
where $\hat{t}_{k,n}\left(b_k\right)= \dfrac{b_k D_{k,n} O}{f_{k,n}^{\rm c}}$ is the computation latency following from \eqref{eq:CompL}, and $T_{\rm s}$ and $T_{\rm ph}$ are the server updating and push latency, respectively. Moreover, we relax the parametric-block lengths (in bits) $\{b_k\}$, to be continuous to simplify the solution of ({P2}), which can be rounded to yield the policy. As mentioned, in large-scale learning models, the values of $\{b_k\}$ are large and the performance loss caused by rounding is negligible. By substituting $t_{k,n}=t$ and relaxing $\{b_k\}$, Problem  ({P2}) is  simplified as
\begin{equation*}~~~~~~~~~~~~~~~~~~~~~~~~~~~~~~(\text{P5})\;
\begin{aligned}
\mathop{\min }\limits_{\{b_k\},t}\; & t \\
{\text{s.t.}}\;\;  
&b_k \geq 0,~\forall k,&(\text{C5.1})\\
&\sum\limits_{k=1}^K b_k = N_{\rm p},&(\text{C5.2})\\
&\sum\limits_{k=1}^K\sum\limits_{n\in \mathcal{G}_k} \rho_{k,n}\left(b_k,t\right)\leq 1,~~~~~~~~~~~~~~~~~~~~~~~~~~~~~~&(\text{C5.3})
\end{aligned}
\end{equation*}
where $\rho_{k,n}\left(b_k,t\right)$ is given in \eqref{eq:Rho}.

\subsubsection{A Useful Property} It is easy to show that ({P5}) is non-convex. To transform the problem into a tractable convex problem, we derive a useful  property. To this end, two definitions are introduced. One is \emph{worker bandwidth allocation rate}, defined as the required bandwidth per additional parameter for one worker. Using \eqref{eq:Rho}, it can be written mathematically as
\begin{equation}\label{eq:WorkerBWRate}
\dfrac{ \partial \rho_{k,n}\left(b_k,t\right)}{\partial b_k} = \dfrac{A_{\rm g}}{  \tilde{t}_{k,n}\left(b_k\right)BR_{{\rm u},k,n} } \left(1+\dfrac{\hat{t}_{k,n}\left(b_k\right) }{\tilde{t}_{k,n}\left(b_k\right)}\right), \;\forall (k,n).
\end{equation}
One can recall that $t$ denotes the one-round latency, $\hat{t}_{k,n}\left(b_k\right)$ and $\tilde{t}_{k,n}\left(b_k\right)$ are the computation latency and uploading time  of worker $(k,n)$, respectively. Note that in \eqref{eq:WorkerBWRate}, the physical meaning of the term, $\dfrac{A_{\rm g}}{\tilde{t}_{k,n}\left(b_k\right)BR_{{\rm u},k,n} }$, is the required bandwidth for worker $(k,n)$ to upload one gradient element; the following scaling factor accounts for computation latency. Next, the \emph{group bandwidth allocation rate} is defined as the required bandwidth per additional parameter for one group, which sums the  bandwidth allocation rates over the workers in the same group as follows: 
\begin{equation}\label{eq:GroupBWRate}
\sum\limits_{n\in \mathcal{G}_k}\dfrac{ \partial \rho_{k,n}\left(b_k,t\right)}{\partial b_k} \overset{\Delta}{=} \sum\limits_{n\in \mathcal{G}_k}\dfrac{A_{\rm g}}{  \tilde{t}_{k,n}\left(b_k\right)BR_{{\rm u},k,n} }\left( 1+\dfrac{\hat{t}_{k,n}\left(b_k\right) }{\tilde{t}_{k,n}\left(b_k\right)} \right),\;1\leq k \leq K.
\end{equation}
From \eqref{eq:WorkerBWRate} and \eqref{eq:GroupBWRate}, two observations can be made. One is that worker/group bandwidth allocation rates depend on both their computation capacities and communication rates. The other is that given fixed one-round latency $t$, the rates  increase as the length of the assigned parametric blocks ($\{b_k\}$) grows. The reason is that heavier load increases computation latency and thereby shortens allowed communication latency, making it necessary to have a larger bandwidth. Based on the above  definitions, one key property of the optimal policy is derived as shown below.
\begin{lemma}[Uniform Group Bandwidth Allocation Rates]\label{thm:OptimumConditions}
\emph{Given a constant $C$, a necessary and sufficient condition for the solution of ({P5}) is 
\begin{equation}\label{eq:UniformGroupBWRates}
\sum\limits_{n\in \mathcal{G}_k}\dfrac{ \partial \rho_{k,n}\left(b_k,t\right) }{\partial b_k} = C, \quad \forall k,
\end{equation}
for a fixed model size,  namely $\sum\nolimits_{k=1}^K b_k = N_{\rm p}$, and  under the constraint on the bandwidth allocation ratios: $\sum\nolimits_{k=1}^K\sum\nolimits_{n\in \mathcal{G}_k} \rho_{k,n}\left(b_k,t\right)= 1$. 
}
\end{lemma}

\proof See appendix \ref{apdx:thmOptimumConditions}.

\subsubsection{Problem of Model Size Maximization}

Though the property in Lemma \ref{thm:OptimumConditions} is insightful, the direct policy computation by  solving the equations in \eqref{eq:UniformGroupBWRates} requires a $(K+2)$-dimensional search, which is impractical when $K$ is large. A more efficient solution can be derived by relating Problem (P5) to the convex problem of model size maximization introduced as follows.

Given one-round latency $t$ for an arbitrary round, let $N_{\rm p}^*\left(t\right)$ denote the maximum size of a model that can be updated within the  round. Then $N_{\rm p}^*\left(t\right)$ solves the following problem of model size maximization 
\begin{equation}\nn \text{(P6)}\qquad \;
 \begin{aligned}
N_{\rm p}^*(t) = \mathop{\max }\limits_{\{b_k\}}\; &\sum\limits_{k=1}^K b_k \\
{\text{s.t.}}\;\;  
&b_k \geq 0,~\forall k,\\
&\sum\limits_{k=1}^K\sum\limits_{n\in \mathcal{G}_k}  \rho_{k,n}\left(b_k,t\right)\leq 1,
\end{aligned}
\end{equation}
where the notation follows that in Problem (P5). {Note that given the one-round latency $t$, if and only if the solution in Problem (P6) satisfies $N_{\rm p}^*(t) \geq N_{\rm p}$, where $N_{\rm p}$ is the target updating model size, Problem (P5) is feasible, as all its constraints can be satisfied.}

 Two useful lemmas for relating Problems ({P5}) and ({P6}) are given as follows, which are proved in Appendices \ref{apdx:lmaRelation} and \ref{apdx:lmaConvexityP6}. 

\begin{lemma}[Relation of maximal feasible model size and latency]\label{lma:Relation}
\emph{
The maximal  model size $N_{\rm p}^*(t)$ is a monotonously increasing function of the one-round latency $t$.
}
\end{lemma}
{It follows from the result in Lemma~\ref{lma:Relation} that the solution for Problem ({P5}) is the minimum latency $t^*$,  for which the updatable model size $N_{\rm p}^*(t^*)$ is no smaller than the target size $N_{\rm p}$. This suggest a solution method  of Problem ({P5}) by a search for $t^*$ using the criterion $N_{\rm p}^*(t^*)\geq N_{\rm p}$ as elaborated in the next subsection. }

This requires solving  Problem (P6) so as to compute   the function $N_{\rm p}^*(t^*)$. To this end, the following result is useful. 

\begin{lemma}\label{lma:ConvexityP6}
\emph{
Given $t$, Problem ({P6}) is convex.
}
\end{lemma}
The convexity allows Problem  ({P6}) to be solved using the traditional primal-dual method. Some needed notations are defined as follows.  Let $\eta_{\lambda}$ and $\{\eta_{b_k}\}$ denote  the step sizes of gradient descent.  The Lagrange function  $\mathcal{L}_{{P6}}\left(\{b_k\},\lambda\right)$ is defined as
\begin{equation}
\mathcal{L}_{{P6}}\left(\{b_k\},\lambda\right) =  - \sum\limits_{k=1}^K b_k+\lambda \left(\sum\limits_{k=1}^K\sum\limits_{n\in \mathcal{G}_k} \rho_{k,n}\left(b_k\right)- 1\right), \text{ with } \lambda\geq 0,
\end{equation}
where $\{\rho_{k,n}\left(b_k\right),\forall (k,n)\}$ are defined in \eqref{eq:RhoT} and $\lambda$ is the multiplier. Using the notations, the application of the  primal-dual method yields   Algorithm \ref{Ag:SolutionP6} for solving Problem (P6). 

\begin{algorithm}
\caption{Model Size Maximization}\label{Ag:SolutionP6}

1: {\bf Input:} $\{R_{{\rm u},k,n}\}$ and $T$.
\begin{itemize}
\item $\{R_{{\rm u},k,n}\}$, the uplink spectrum efficiencies of all workers.
\item $T$, the given one-round latency.
\end{itemize}

2: {\bf Initialize} $t=T$, $\lambda^{(0)}$, and $l=0$.

3: {\bf Loop}

4: \;\;\;\;$\lambda^{(l+1)} = \max\left\{\lambda^{(l)} +\eta_{\lambda} \dfrac{\partial \mathcal{L}_{{P6}}\left(\{b_k\},\lambda\right) }{\partial \lambda}, 0\right\}$.

5: \;\;\;\;{\bf Initialize} $\{b_k^{(0)}\}$ and $i=0$.

6: \;\;\;\;{\bf Loop}

7: \;\;\;\;\;\;\;\;$b_k^{(i+1)} = b_k^{(i)} - \eta_{b_k}\left( -1 + \lambda^{(l+1)} \sum\limits_{n\in \mathcal{G}_k} \dfrac{\partial \rho_{k,n}\left(b_k\right) }{\partial b_k}\right),\;\forall k$.

8: \;\;\;\;\;\;\;\;$i=i+1$.

9: \;\;\;\;{\bf Until Convergence}

10:\;\;\;\;$\{b_k^* = b_k^{(i-1)},\forall k\}$ and $l=l+1$.

11:{\bf Until Convergence}

12:Get $N_{\rm p}^*(T)=\sum\nolimits_{k=1}^K b_k^*$.

13:{\bf Output:} $N_{\rm p}^*(T)$ and $\{b_k^*\}$.

\end{algorithm}

\subsubsection{Solution by Nested Optimization} 

It follows from the results in the preceding subsection that Problem (P5) can be solved by nesting a one-dimensional search over $t$  and the solution of a convex problem, namely Problem (P6). The search can be efficiently implemented using the bisection method given the monotonicity in Lemma \ref{lma:Relation} while the solution of Problem (P6) relies on Algorithm \ref{Ag:SolutionP6}.  Nesting them yields Algorithm \ref{Ag:JointOptimalDesign} for computing the optimal joint PABA policy. 

{ The computational complexity of Algorithm \ref{Ag:JointOptimalDesign} can be divided into two parts. One is the inner loop for solving the convex Problem (P6). Its complexity is $\mathcal{O}(K^3)$, where $K$ is the number of worker groups. The other is the outer loop of bisection search, with complexity of $\mathcal{O}\left[\log(1/\delta)\right]$ and $\delta$ being the convergence tolerance. The overall computational complexity of Algorithm \ref{Ag:JointOptimalDesign} is $\mathcal{O}\left[\log(1/\delta)K^3\right]$. As $K$ is small (e.g., $K=15$) compared with the number of parameters (e.g., $N_{\rm p}>10^6$), the computational overhead caused by Algorithm 4 at the server can be ignored.}

\begin{algorithm}
\caption{Optimal Joint Parameter Allocation and Bandwidth Allocation}\label{Ag:JointOptimalDesign}
1: {\bf Input:} $\{R_{{\rm u},k,n}\}$.
\begin{itemize}
\item $\{R_{{\rm u},k,n}\}$, the uplink spectrum efficiencies of all workers.
\end{itemize}

2: {\bf Select} $t_{\rm u} = T_{\rm u}$ so that $t=t_{\rm u}$ makes $N_{\rm p}^*(t_{\rm u})$  defined in ({P6}) larger than $N_{\rm p}$.

3:  {\bf Select} $t_{\rm l} = T_{\rm l}$ so that $t=t_{\rm l}$ makes $N_{\rm p}^*(t_{\rm l})<N_{\rm p}$.

4: {\bf While} $t_{\rm u }\not= t_{\rm l}$

4: \;\;\;\;Let $t_{\rm m} = (t_{\rm u}+t_{\rm l})/2$. And substitute $t=t_ {\rm m}$ in to ({P6}).

5: \;\;\;\;Solve ({P6}) with Algorithm \ref{Ag:SolutionP6} to obtain $N_{\rm p}^*(t_{\rm m})$ and $\{b_k^*\}$ by inputing $\{R_{{\rm u},k,n}\}$ and $t_{\rm m}$.

6: \;\;\;\;{\bf If} $N_{\rm p}^*(t_{\rm m})\geq N_{\rm p}$

7: \;\;\;\;\;\;\;\;$t_{\rm u} =  t_{\rm m}$.

8: \;\;\;\;{\bf Else}

9: \;\;\;\;\;\;\;\;$t_{\rm l} = t_{\rm m}$.

10:\;\;\;\;{\bf End if}

11:{\bf End while}

12:$t^* = t_{\rm m}$.

13:{\bf Output:} $t^*$ and $\{b_k^*\}$.
\end{algorithm}

\subsection{Two Special Cases}

\subsubsection{Single-Worker Groups}
Consider the  case when each group comprises only a single worker. The dataset is not partitioned and each worker uses the whole dataset to update an assigned parametric block. Using the property of uniform group allocation rates in \eqref{eq:UniformGroupBWRates}, the parametric-block length can be derived as functions of the latency $t$ and the unknown variable $C$:
\begin{equation}\label{eq:bk_OneWorker}
b_k\left(t,C\right)= \dfrac{f_{k,1}}{D_{k,1}O}\left(t-T_{\rm ph}-T_{\rm s} - \sqrt{ \dfrac{A_{\rm g}t}{  CR_{{\rm u},k,1} }  }\right), \forall k.
\end{equation}
From \eqref{eq:bk_OneWorker}, the number of parameters assigned to group $\mathcal{G}_k$ decreases with its computation time to calculate one gradient element, say  $\dfrac{f_{k,1}}{D_{k,1}O}$, and increases with its uploading rate $R_{k,1}^{\rm u}$. This is aligned with the analysis in \eqref{eq:RBS}.
By substituting $b_k\left(t,C\right)$ in \eqref{eq:bk_OneWorker} into the constraints (C5.2) and (C5.3), it's easy to show that the following equations hold  if  the latency is at its minimum: 
\begin{equation}\label{eq:Case1}
\left\{
\begin{aligned}
&\sum\limits_{k=1}^K b_k \left(t^*, C\right) = N_{\rm p},\\
&\sum\limits_{k=1}^K\sum\limits_{n\in \mathcal{G}_k} \rho_{k,n}\left(b_k\left(t^*, C\right)\right)= 1.
\end{aligned}
\right.
\end{equation}
Using \eqref{eq:Case1}, the optimal latency $t^*$ and the corresponding $C$ can be efficiently solved for since  there are only  two variables. Then the optimal parameter allocation  $\{b_k^*\}$ follows from \eqref{eq:bk_OneWorker}.  Substituting $\left\{\{b_k^*\}, t^*\right\}$ into 
\eqref{eq:BWS} gives  the optimal bandwidth allocation  $\{\rho_{k,1}^*\}$. 

\subsubsection{Uniform Intra-group  Computation Capacities} In this case, all workers within the same group, say group $\mathcal{G}_k$, have identical computation capacities and hence the same computation time: $\left\{\dfrac{D_{k,n}}{f_{k,n}^{\rm c}} = \dfrac{D_{k,1}}{f_{k,1}},\;\forall n\in \mathcal{G}_k\right\}$.
In this case, a similar load-balancing scheme as the preceding special  case can be derived: 
\begin{equation}
b_k \left(t,C\right)= \dfrac{f_{k,1}}{D_{k,1}O}\left(t-T_{\rm ph}-T_{\rm s} - \sqrt{ \dfrac{A_{\rm g}t}{  C } \sum\limits_{n\in \mathcal{G}_k}\dfrac{1}{R_{{\rm u},k,n}} }\right), \forall k.
\end{equation}
A similar solution method can be applied to compute the optimal PABA policy, specified by  the optimal  load assignments $b_k \left(t^*,C\right)$ and  bandwidth allocation  $\{\rho_{k,n}^*\}$.

\section{Experimental  Results}
\subsection{Experiment Setup} 
Consider a single-cell wireless network in a disk with a  radius of $0.15$ kilometres. The AP (edge {server}) is located at the center with multiple  workers randomly located within the disk. The workers are separated in to $K$ groups each having $N$ workers. The total bandwidth is $B$. By default, their values are set as $K=15$, $N=15$, and $B= 100$ MHz unless specified otherwise. The workers' computation capacities are uniformly selected from the set $\{0.1, 0.2, \cdots, 1.0\}$ GHz. The learning task is a $\ell_1$-regularized logistic regression task for training a news-filtering model using  the News20 dataset collected in \cite{lang1995newsweeder}. The model has $N_{\rm p} = 1,241,220$ parameters. The training dataset contains $15936$ samples and the test dataset contains $3993$ samples. For each group, the training dataset is uniformly partitioned into $N$ subsets. Each subset is downloaded by one worker. Wireless channels are modelled with the following parameters. The noise power density is $N_0 = {\rm{-174\; dBm/Hz}}$. The transmission power of AP and workers is $P_{\rm b} = 46\; {\rm dBm}$ and $P_{\rm u} = 24\; {\rm dBm}$, respectively.  The path loss between worker and AP is $128.1+37.6\log d $ with the distance $d$ in kilometre. Rayleigh fading is assumed.

{
For comparison, four algorithms are considered, described as follows \footnote{The source codes for implementing PARTEL are available at {https://www.eee.hku.hk/\%7ewirelesslab/resources/Demo.zip}. }.
\begin{itemize}
\item \emph{Baseline}: The number of parameters assigned to one worker is proportional to its computation capability and bandwidths are equally allocated. 

\item \emph{Bandwidth-aware parameter allocation}: The bandwidth is first equally allocated and then the parameters are allocated by the scheme of bandwidth aware parameter allocation in Algorithm \ref{Ag:LLB}.

\item \emph{Parameter-aware bandwidth allocation}: The parameters are proportionally allocated first and then the bandwidth is allocated using the scheme of parameter aware bandwidth allocation in Algorithm \ref{Ag:BWS}.

\item \emph{Joint PABA}: The parameter allocation and bandwidth allocation are jointly optimized using Algorithm 4.
\end{itemize}
}


\subsection{Learning Performance} 
\begin{figure}[t]
    \centering
   \begin{subfigure}[b]{0.49\textwidth}
       \includegraphics[width=\textwidth]{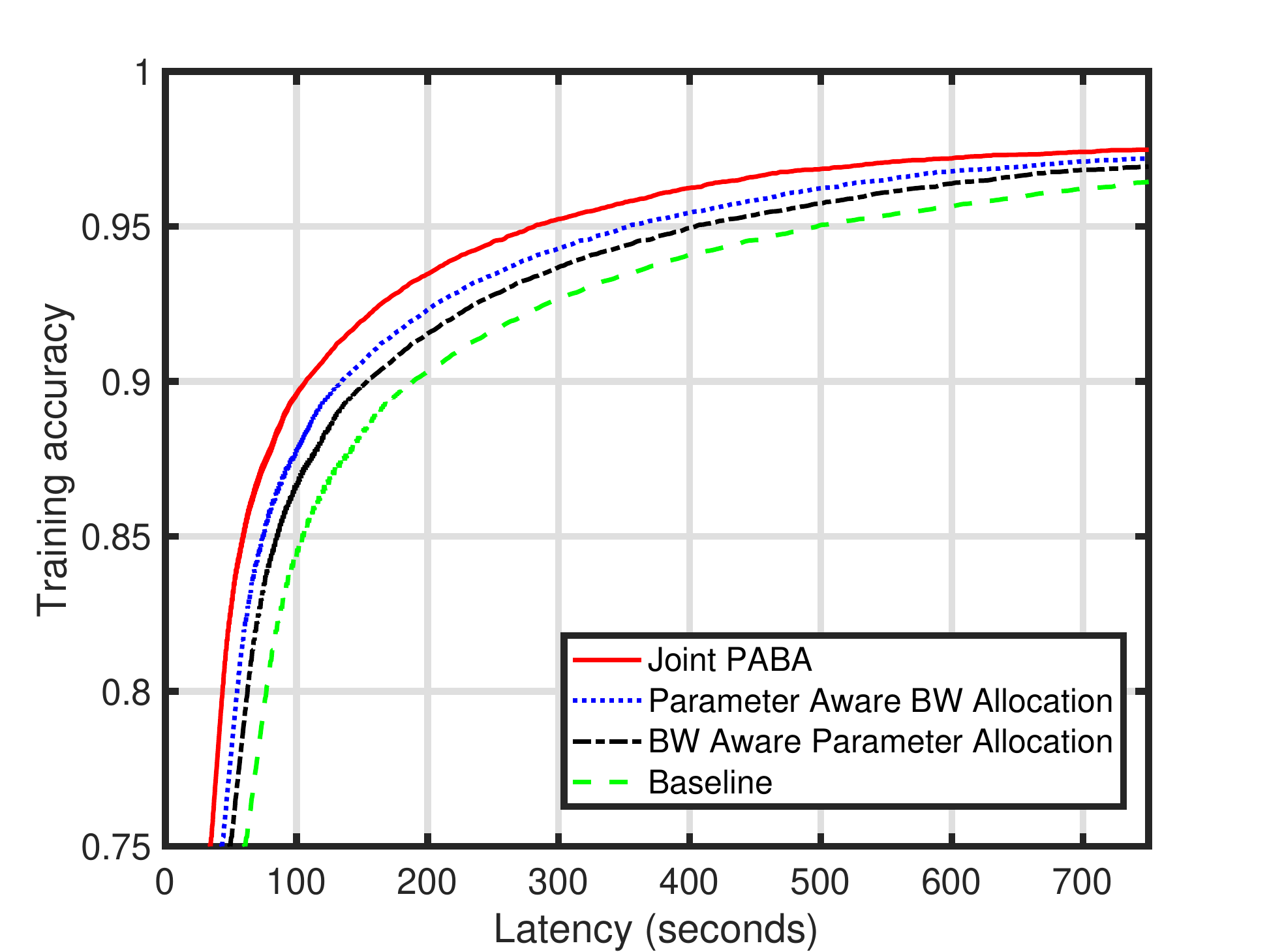}
        \caption{Training accuracy versus latency.}
    \end{subfigure}
    \begin{subfigure}[b]{0.49\textwidth}
        \includegraphics[width=\textwidth]{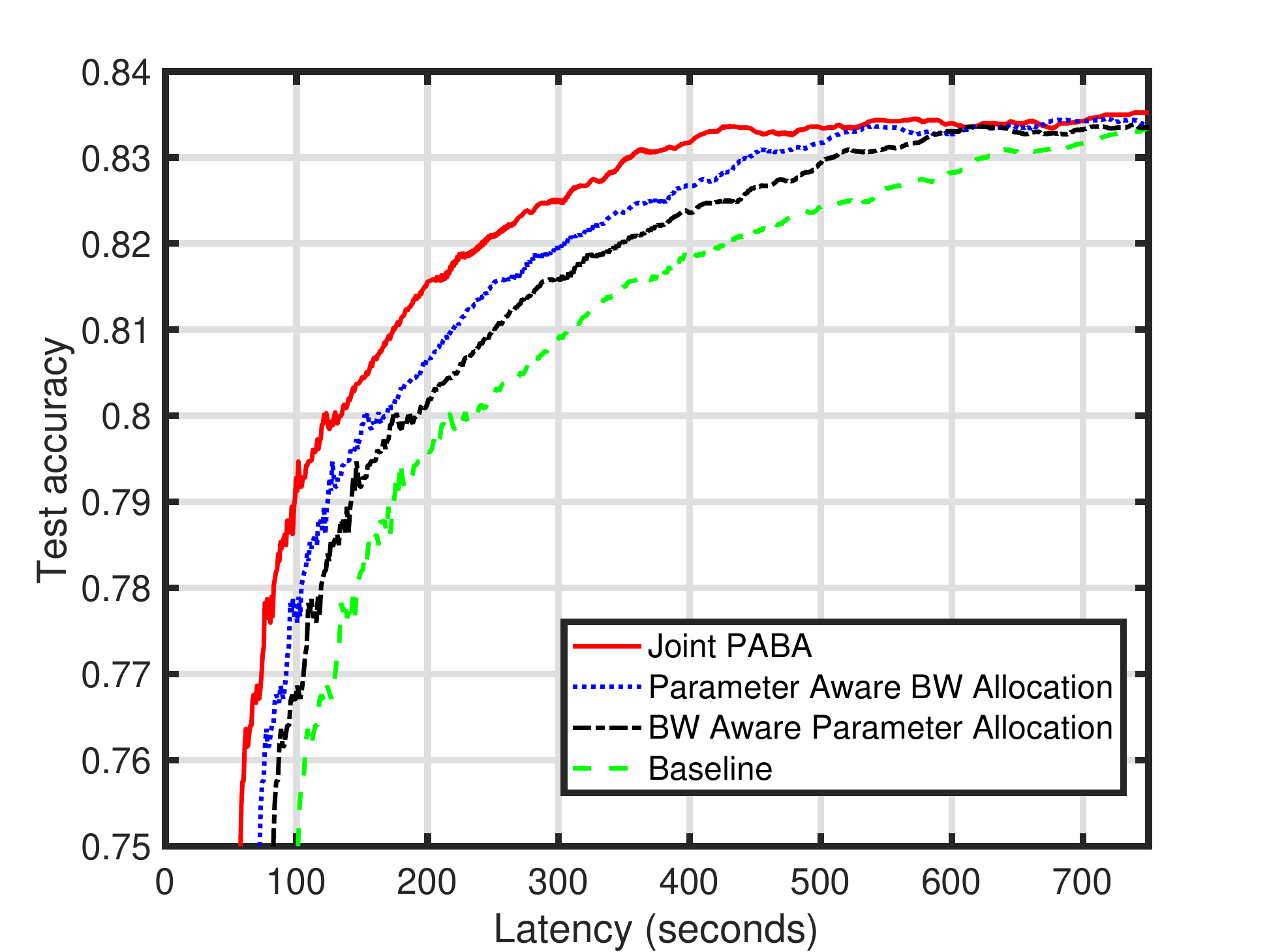}
        \caption{Test accuracy versus latency.}
    \end{subfigure}
    \caption{Learning performance  versus (communication-plus-computation) latency.}\label{fig:TrainingTime}
\end{figure}

Latency minimization PABA can accelerate  the model  convergence. To evaluate the gain, the curves of (model) training and test accuracies versus latency are plotted in Fig. \ref{fig:TrainingTime}. As observed, the partially integrated and joint  PABA algorithms outperform the baseline algorithm in terms of model convergence. For example, given the latency of $100$ second,  the proposed joint PABA,  parameter aware bandwidth allocation, and bandwidth aware parameter allocation achieve a training accuracy of (5.24\%, 3.49\%, 2.26\%) and a test accuracy of (4.40\%, 2.84\%, 1.99\%) higher than that of the baseline algorithm, respectively. {With respect to the baseline, the performance gains of the PABA schemes are due to the integration of workload and bandwidth. Specifically, additional bandwidth (or less workload) is allocated to a device to compensate for slow computation (or weak link), thereby reducing the per-round latency. The PABA schemes with varying complexity support different degrees of integration and as a result, achieve different levels of latency reduction with joint PABA performing the best.}



{
\subsection{PARTEL v.s. FEEL}
In Fig. \ref{fig:CMP}, the proposed PARTEL framework and the FEEL framework are compared. As mentioned in Remark \ref{Rmk:RelationFEEL}, the PARTEL framework reduces to the FEEL framework in the case of only one group. Consider there are 50 workers in the cell. For PARTEL, they are clustered into 5 groups, each of which has 10 workers. For FEEL, they are in one group and hence no model partition. For fairness, we use the same scheme of joint PABA to compare PARTEL and FEEL. From the figure, the proposed PARTEL framework outperforms the FEEL framework with a latency reduction of 48.43\% on average to achieve the same accuracy. The reason is as follows. For the PARTEL framework, each worker only needs to transmit the gradient of a parametric block while for the FEEL framework, the gradient of the whole parameter vector is needed to be uploaded by each worker. This makes the uploading latency much larger for the latter, especially when the wireless resources are limited.

\begin{figure}[t]
    \centering
   \begin{subfigure}[b]{0.49\textwidth}
       \includegraphics[width=\textwidth]{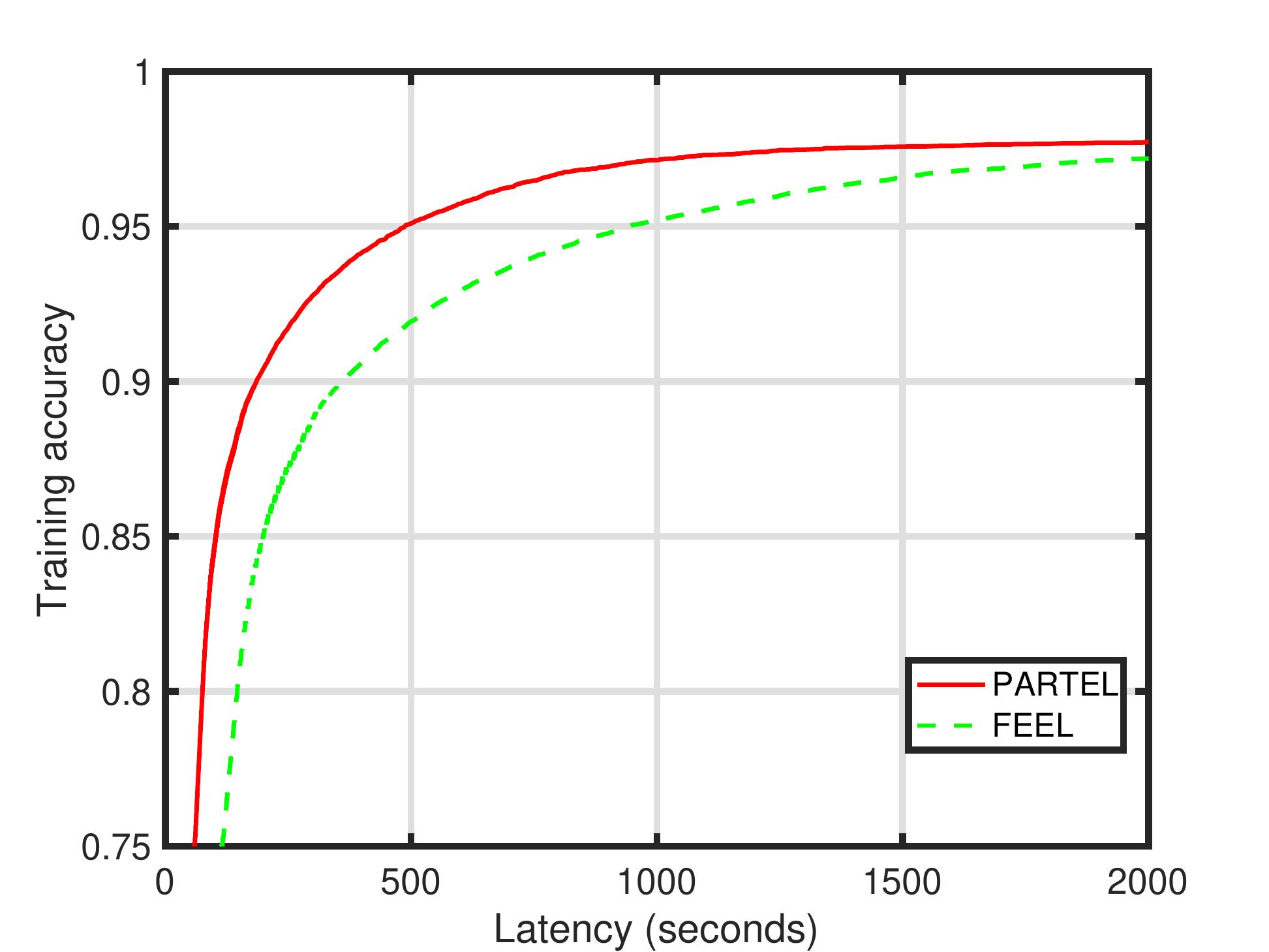}
        \caption{Training accuracy versus latency.}
    \end{subfigure}
    \begin{subfigure}[b]{0.49\textwidth}
        \includegraphics[width=\textwidth]{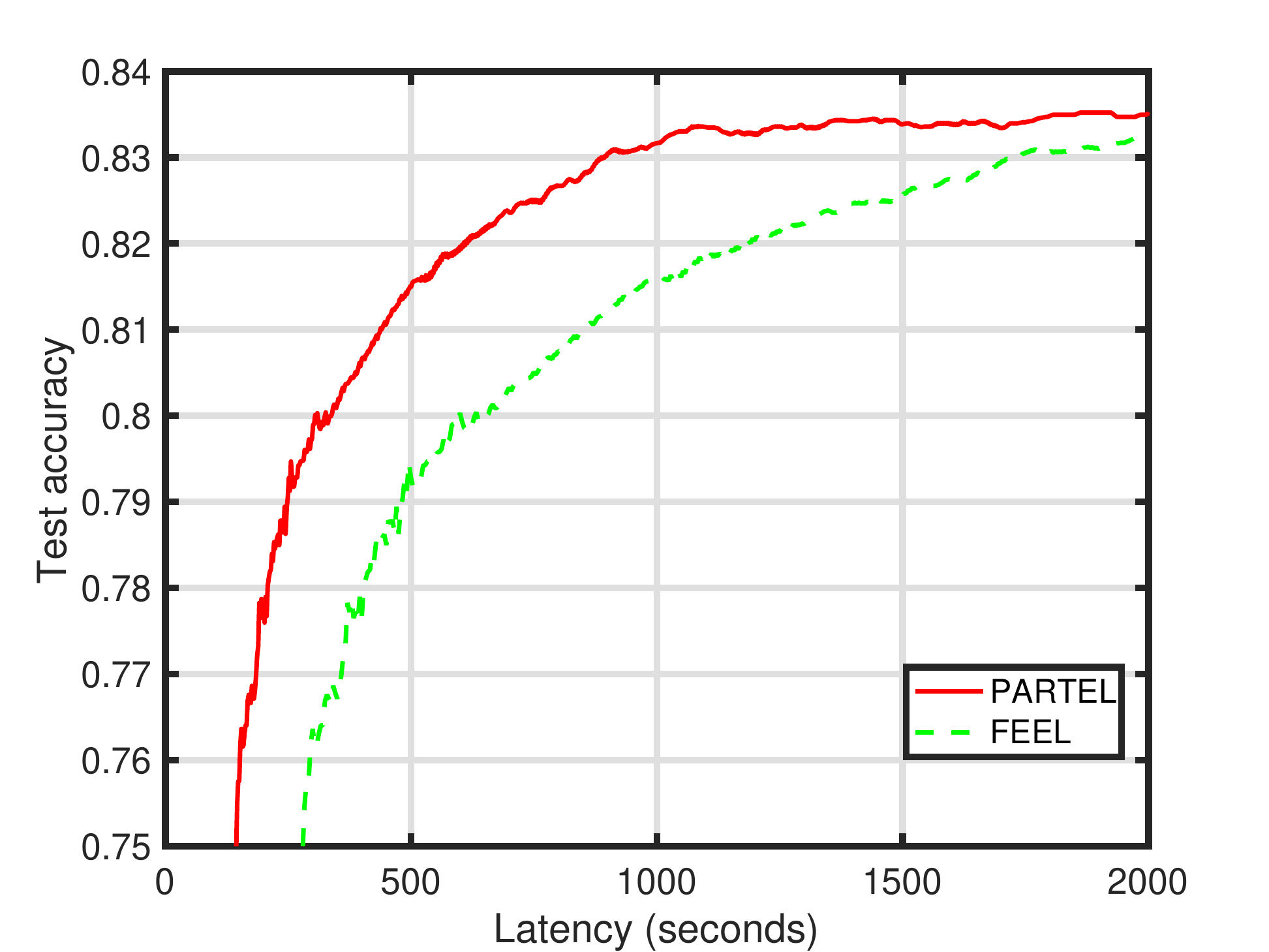}
        \caption{Test accuracy versus latency.}
    \end{subfigure}
    \caption{Comparison between PARTEL and FEEL.}\label{fig:CMP}
\end{figure}
}

\subsection{Latency Performance}

\begin{figure}[t]
    \centering
   \begin{subfigure}[b]{0.49\textwidth}
       \includegraphics[width=\textwidth]{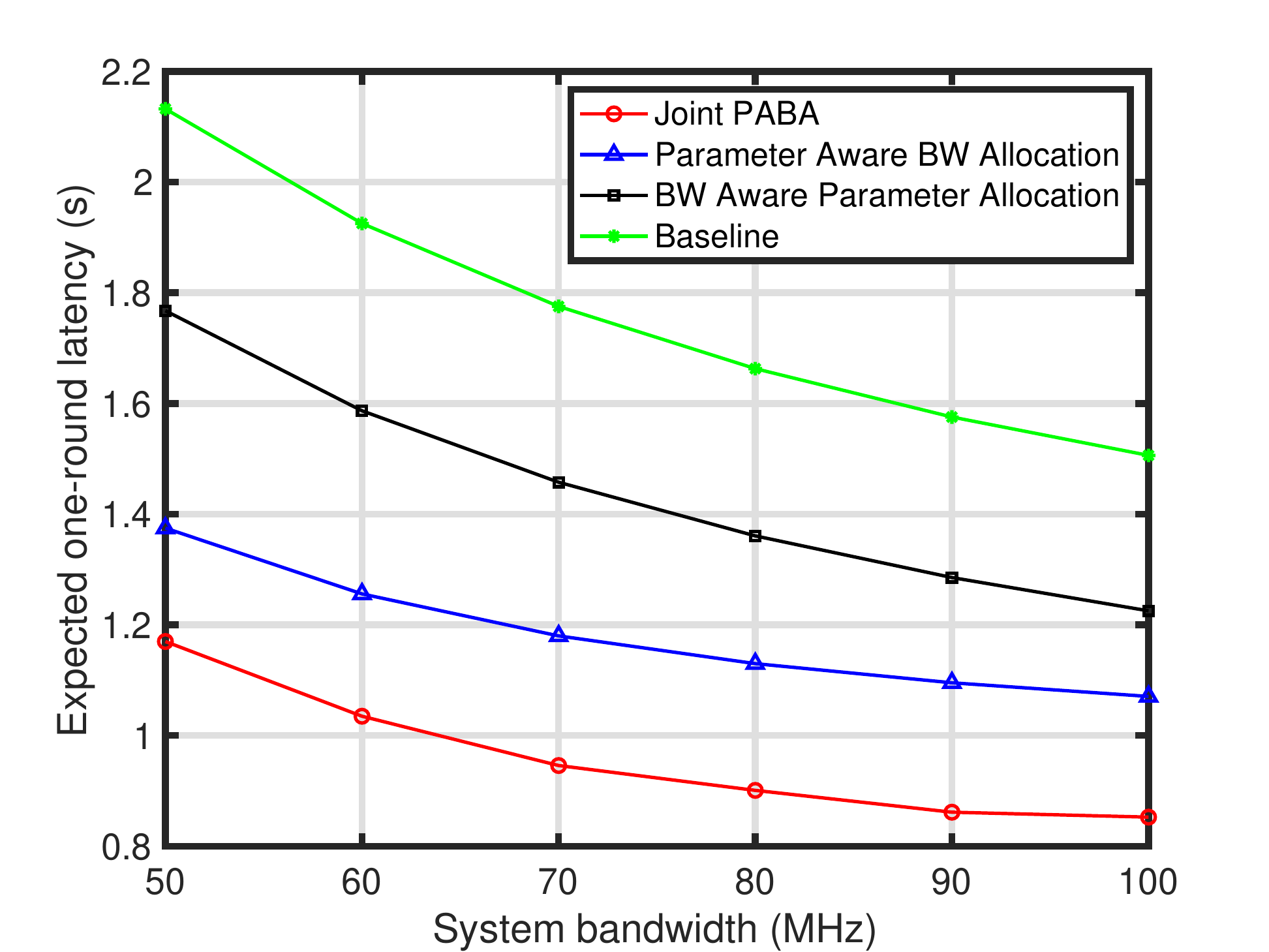}
        \caption{Effect of Bandwidth}
    \end{subfigure}
    \begin{subfigure}[b]{0.49\textwidth}
        \includegraphics[width=\textwidth]{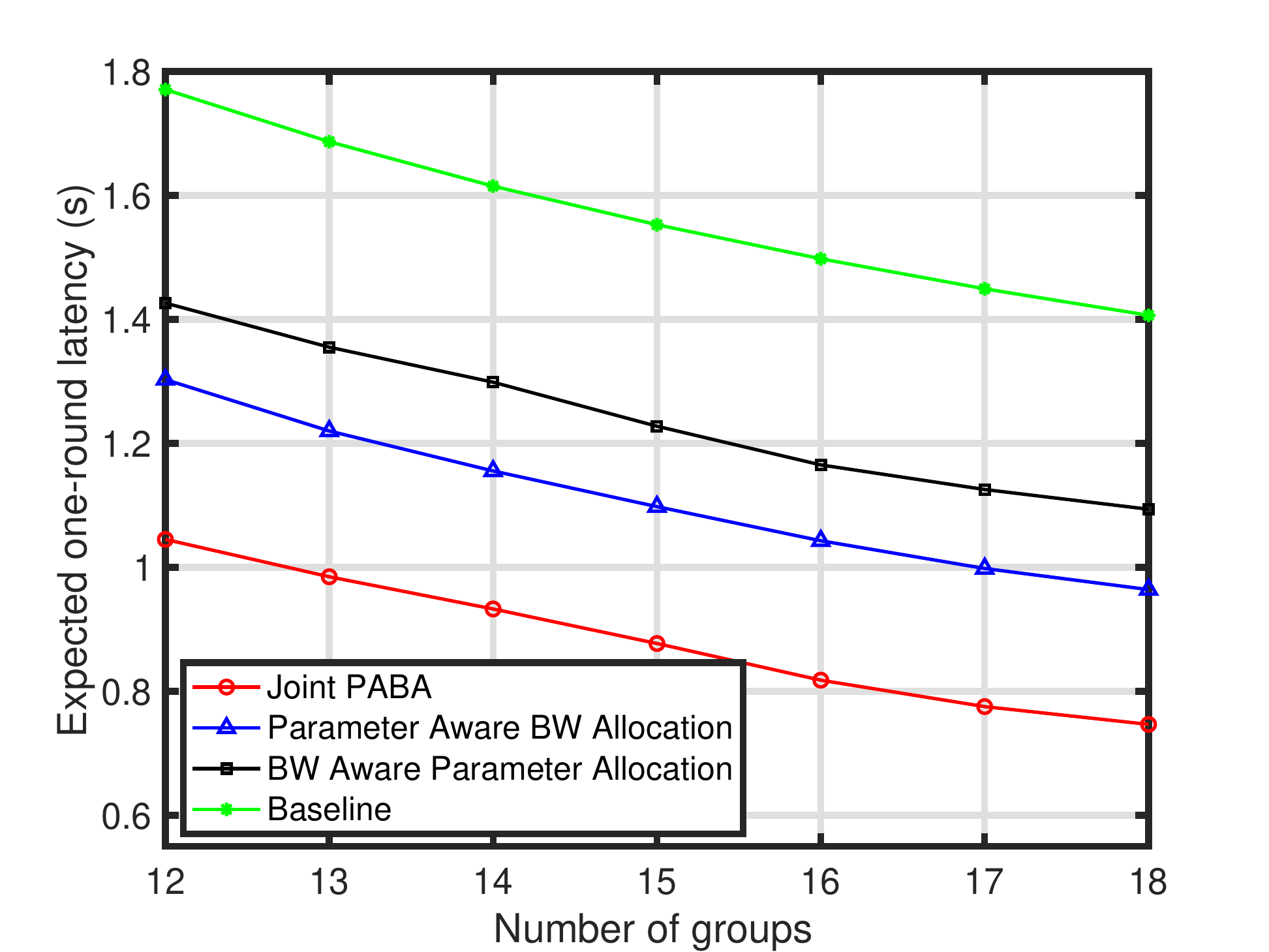}
                \caption{Effect of Number of Groups}
    \end{subfigure}
    \caption{Latency performance comparison  for (a) a varying bandwidth and (b) a varying number of worker groups. }\label{fig:Bandwidth}
\end{figure}

The latency performance of joint and partially integrated PABA and baseline scheme in terms of expected one-round latency are compared in Fig.~\ref{fig:Bandwidth} for a varying bandwidth and a varying number for worker groups. First, as expected, the latency of all algorithms are observed to decrease as either the bandwidth or group number increase, representing more communication and computation resources, respectively. For a large bandwidth (or a  group number), the latency saturates as it is dominated by computation latency (or communication latency). Next, the PABA algorithms are observed to significantly reduce the latency with respect to the baseline scheme. In particular, joint PABA achieves latency reduction of 46.73\% for the bandwidth of 70 MHz and 46.92\% for the number of groups equal to 18. Among the PABA algorithms, joint PABA outperforms two partially integrated PABA algorithms at the cost of higher complexity. On the other hand, the latency comparison between parameter aware bandwidth allocation and bandwidth aware parameter allocation  suggests the former is more effective. Because the former can cope with channel heterogeneity for all workers while the latter can only cope with the computation capacity heterogeneity in worker group level. 

\begin{figure}[h]
\centering
\includegraphics[width=0.49\textwidth]{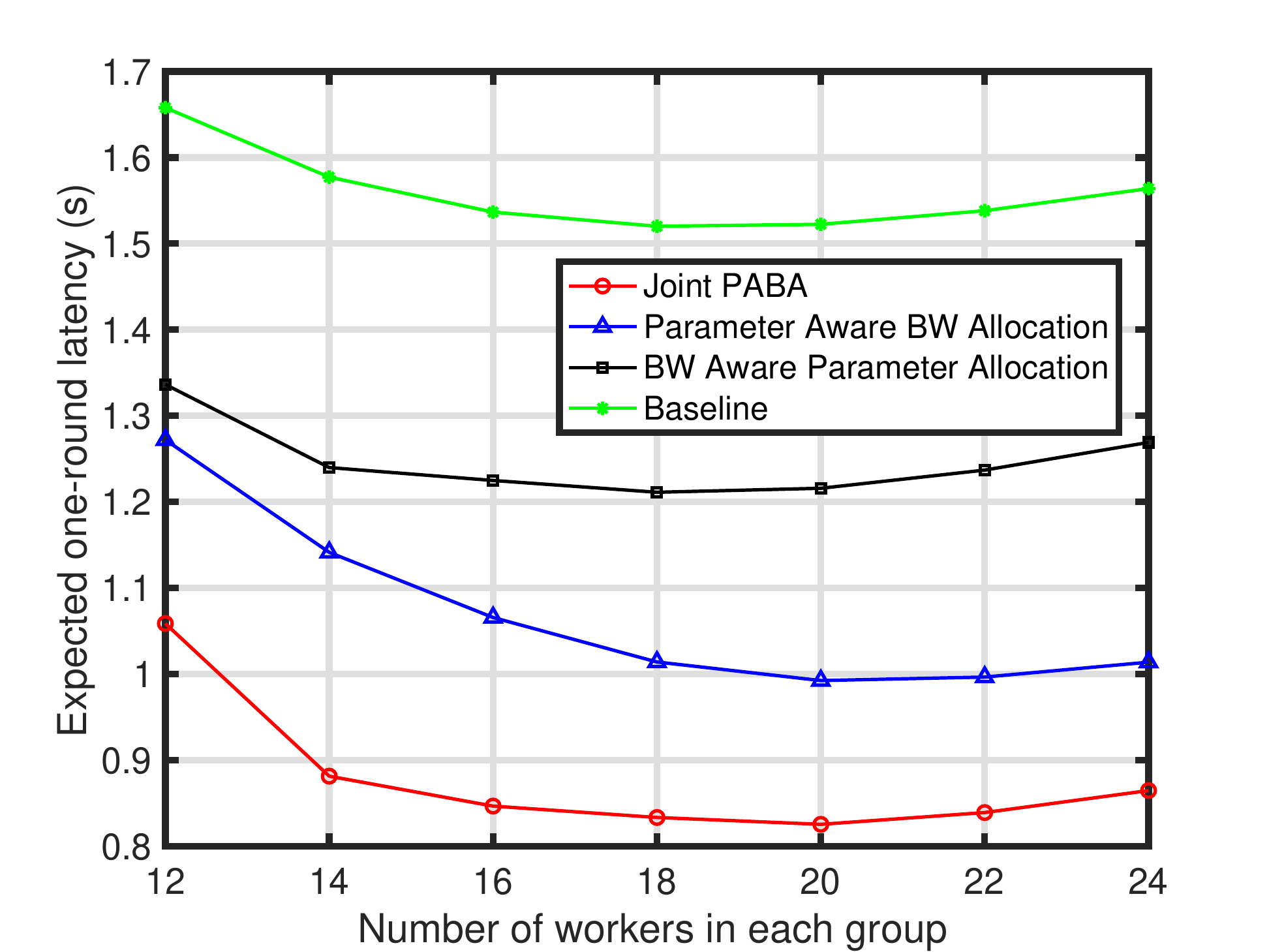}
\caption{One-round latency versus number of workers in each group}\label{fig:Worker}
\end{figure}

Consider the case where the number of worker groups is fixed but the group size grows. The growth has two conflicting effects. On one hand, the computation load, say the number of assigned parameters, per worker  reduces, resulting in decreasing  computation latency. On the other hand, more workers sharing a fixed bandwidth causes increases the  communication latency. This suggests an optimal group size  for learning latency minimization as confirmed by the curves of latency versus number of workers per group plotted in Fig. \ref{fig:Worker}. The optimal group size differs for different algorithms e.g., 20 for joint PABA and 18 for the baseline scheme.

The simulation results above show that the proposed joint PABA scheme has the best performance and verify our analysis in Sections IV, V, and VI.

\section{Conclusion}
In this paper, we have proposed the new edge-learning framework, PARTEL, for performing a large-scale learning task in a wireless network. The framework features both data-and-model partitioning for distributing learning  at many resource-constrained mobile devices. For efficient edge implementation of PARTEL, we have jointly designed the functional blocks of \emph{parameter allocation and bandwidth allocation} (PABA), resulting in substantial latency reduction. 

The current work opens several interesting directions for future investigation. One direction is to extend the joint PABA design to {the cases of multi-cell systems with inter-cell interference management required and coexisting of networks/services (PARTEL and non-PARTEL users).}
 Another direction is to investigate worker scheduling to balance distribute computation capacities and multi-access latency. 
Designing communication techniques for  PARTEL such as multi-antenna and millimeter-wave transmission is also an interesting direction to explore. 

\appendix
\subsection{Proof of Lemma \ref{lma:NIter}}\label{apdx:NIter}
In the case of smooth regularization, both $\mathcal{F}({\bm \theta})$ and $\mathcal{R}({\bm \theta})$ are smooth functions. For an arbitrary round, say the $(r+1)$-th, gradient descent is used to update the model parameters:
\begin{equation}
\theta_{n_{\rm p}}^{(r+1)} = \theta_{n_{\rm p}}^{(r)} + \eta_r \nabla_{\theta_{n_{\rm p}}^{(r)}} \mathcal{L}\left({\bm \theta}^{(r)} \right), 1 \leq n_{\rm p} \leq N_{\rm p},
\end{equation}
where $\theta_{n_{\rm p}}^{(r)}$ is the ${n_{\rm p}}$-th element of ${\bm \theta}^{(r)}$, $\eta_r$ is the learning rate, and $\nabla_{\theta_{n_{\rm p}}^{(r)}} \mathcal{L}\left({\bm \theta}^{(r)} \right)$ is the gradient defined in \eqref{eq:BlockGradient}. Besides, due to the decomposable structure of the objective function $\mathcal{L}(\cdot)$, 
$\nabla_{\theta_{n_{\rm p}}^{(r)}} \mathcal{L}\left({\bm \theta}^{(r)} \right)$ is independent of $\nabla_{\theta_{n_{\rm p}}^{'(r)}} \mathcal{L}\left({\bm \theta}^{(r)} \right)$, for any $n_{\rm p} \not=n_{\rm p}^{'}$. Hence, in the distributed algorithms, the ground-true gradients can be calculated for each parametric block. Thereby, the distributed gradient descent algorithm is equivalent to the centralized one.
 
In the case of non-smooth regularization, $\mathcal{R}({\bm \theta})$ is non-smooth. For an arbitrary round, say the $(r+1)$-th, proximal gradient descent is used to update the learning parameters:
\begin{equation}
{\bm \theta}^{(r+1)} = {\rm prox}\left({\bm \theta}^{(r)} - \eta_r \nabla \mathcal{F}({\bm \theta}^{(r)})\right),
\end{equation}
where ${\rm prox}({\bf y}) = \arg\min\limits_{\bm \mu} \big( \mathcal{R}(\bm \mu) + \dfrac{1}{2}||\bm \mu-{\bf y}||_2^2 \big)$. Similarly,  for any $n_{\rm p}\not=n_{\rm p}^{'}$, $ \nabla_{\theta_{n_{\rm p}} }\mathcal{F}$ is independent of $\nabla_{\theta_{n_{\rm p}^{'}} } \mathcal{F}$. Besides, $\mathcal{R}(\bm \mu)$ and $||\bm \mu-{\bf y}||_2^2$ are block separable. Thereby, the distributed proximal gradient descent algorithm is equivalent to the centralized one.

In summary, the distributed implementation has no impact on the calculated proximal gradients or  gradients. Hence, the convergence rates (in rounds) only depend on the learning algorithms themselves and are irrelevant to parameter allocation and bandwidth allocation. 

\subsection{Proof of Theorem \ref{thm:RBS}}\label{apdx:RBS}
After relaxation, the first constraint of \eqref{eq:MILP} turns to be $\{b_k\geq 0, \; 1\leq k \leq K\}$, while the other parts remain the same. The relaxed problem is a linear program. 
KKT conditions are used to solve the linear program. The Lagrange function can be written as
\begin{equation}
\mathcal{L}\left(\{b_k\},t_{\rm PA},\{\lambda_k\}\right)=t_{\rm PA} + \mu \left(\sum\limits_{k=1}^K b_k - N_{\rm p}\right) + \sum\limits_{k=1}^K \lambda_k\left(t_k\left(b_k\right)-t_{\rm PA}\right),
\end{equation}
where $\mu$ and $\{\lambda_k\}$ are the Lagrangian multipliers. The KKT conditions are given by
\begin{equation}\label{eq:apKKTLP}
\left\{
\begin{aligned}
&\dfrac{\partial \mathcal{L}\left(\{b_k\},t_{\rm PA},\{\lambda_k\}\right)}{\partial b_k} = \mu+\lambda_k \max\limits_{n\in \mathcal{G}_k}\left\{\dfrac{D_{k,n} O}{f_{k,n}^{\rm c}}+ \dfrac{ A_{\rm g} }{\rho_{k,n}^*BR_{{\rm u},k,n}}\right\} = 0,\;1\leq k\leq K,\\
&\dfrac{\partial \mathcal{L}\left(\{b_k\},t_{\rm PA},\{\lambda_k\}\right)}{\partial t_{\rm PA}} = 1-\sum\limits_{k=1}^K \lambda_k=0,\\
&\lambda_k\left(t_k\left(b_k\right)-t_{\rm PA}\right)=0,\;1\leq k\leq K,\\
&\sum\limits_{k=1}^K b_k -  N_ {\rm p}=0, \lambda_k\geq 0,\; t_k\left(b_k\right)-t_{\rm PA}\leq 0,\;1\leq k\leq K,\\
\end{aligned}
\right.
\end{equation}
In \eqref{eq:apKKTLP}, the second condition indicates $\exists k,\; \lambda_k\neq 0$. Then, together with the first condition, 
we can show $\mu\neq0$, and further show that $\{\lambda_k\neq0,\forall k\}$ as $\mu\neq0$. Next, from the third condition, we have 
\begin{equation}\label{eq:apNC}
t_k\left(b_k\right)-t_{\rm PA}=0,\;1\leq k\leq K.
\end{equation}
By substituting the group latency $t_k\left(b_k\right)$ in \eqref{eq:GroupLinear} into the above equation,  \eqref{eq:OPBSL} can be derived.
 Next, by substituting \eqref{eq:OPBSL} into the constraint $\sum\nolimits_{k=1}^K b_k = N_{\rm p}$, we can get \eqref{eq:RBS}.
In \eqref{eq:RBS}, the optimum $t_{\rm PA}^*$ can be solved by bisection search because $N_{\rm p}$ increases with $t_{\rm PA}$. Then, the optimal parameter-allocation scheme $\{b_k^*\}$ is given in \eqref{eq:OPBSL}. 

\subsection{Proof of Lemma \ref{lma:CBA}}\label{apdx:CBA}
({P4}) is convex if its objective function is convex, as the constraint is a linear set.
First, for any one worker, say worker $(k,n)$, by substituting the push latency in \eqref{eq:PushLa}, the computation latency in \eqref{eq:CompL}, and the pull latency in \eqref{eq:PullL}, its total latency in \eqref{eq:WorkerL} can be derived as
\begin{equation}\label{eq:apWL}
\begin{aligned}
t_{k,n} \left(\rho_{k,n}\right) &= T_{\rm ph} + \hat{T}_{k,n} + \tilde{t}_{k,n}\left(\rho_{k,n}\right)+T_{\rm s}=T_{\rm ph} + \hat{T}_{k,n} + \dfrac{\hat{b}_k^* A_{\rm g} }{\rho_{k,n}BR_{{\rm u},k,n}}+T_{\rm s},
\end{aligned}
\end{equation}
where $T_{\rm ph}$, $\hat{T}_{k,n} = \hat{t}_{k,n}(\hat{b}_k^*)$, and $T_{\rm s}$ are constants. In \eqref{eq:apWL}, $t_{k,n} \left(\rho_{k,n}\right)$ is a convex function of $\rho_{k,n}$. Then, the objective function of  ({P4}), given by $\max\nolimits_k t_k\left(\{\rho_{k,n}\}\right) = \max\nolimits_{(k,n)}t_{k,n}\left(\rho_{k,n}\right)$,
is also convex, because max operation preserves convexity. 

\subsection{Proof of Theorem \ref{thm:BWS}}\label{apdx:BWS}
To solve ({P4}), we derive and solve an equivalent convex problem. First, define $t_{\rm BA}=\mathop{\max }\nolimits_k  t_k\left(\{\rho_{k,n}\}\right)$. 
Then, substituting it and $t_k\left(\{\rho_{k,n}\}\right) = \max_{n\in \mathcal{G}_k} t_{k,n}\left(\rho_{k,n}\right)$ into ({P4}), it can be equally derived as
\begin{equation}\label{eq:apEQP4}
\mathop{\min }\limits_{\{\rho_{k,n}\},t_{\rm BA}}\;  t_{\rm BA}, \; {\text{s.t.}}\; \sum\limits_{k=1}^K\sum\limits_{n\in \mathcal{G}_k} \rho_{k,n} \leq 1, \; \& \; t_{k,n}\left(\rho_{k,n}\right)\leq t_{\rm BA},~\forall (k,n).
\end{equation}
KKT conditions are used to solve the convex problem in \eqref{eq:apEQP4}. The Lagrange function is
\begin{equation}
\mathcal{L}\left(\{\rho_{k,n}\},t_{\rm BA},\{\lambda_{k,n}\}\right) = t_{\rm BA}+\mu\left( \sum\limits_{k=1}^K\sum\limits_{n\in \mathcal{G}_k} \rho_{k,n} - 1 \right) + \sum\limits_{k=1}^K\sum\limits_{n\in \mathcal{G}_k} \lambda_{k,n} \left( t_{k,n}\left(\rho_{k,n}\right)- t_{\rm BA} \right),
\end{equation}
where $\mu$ and $\{\lambda_{k,n}\}$ are the multipliers, $t_{k,n}\left(\rho_{k,n}\right)$ in \eqref{eq:apWL}. Then, the KKT conditions are 
\begin{equation}\label{eq:apKKTCP}
\left\{
\begin{aligned}
&\dfrac{\partial \mathcal{L}\left(\{\rho_{k,n}\},t_{\rm BA},\{\lambda_{k,n}\}\right)}{\partial \rho_{k,n}} = \mu - \lambda_{k,n}  \dfrac{ \hat{b}^*_k A_{\rm g} }{\rho_{k,n}^2B R_{{\rm u},k,n}} = 0,~\forall (k,n),\\
&\dfrac{\partial \mathcal{L}\left(\{\rho_{k,n}\},t_{\rm BA},\{\lambda_{k,n}\}\right)}{\partial t_{\rm BA}} = 1 - \sum_{k=1}^K\sum_{n\in \mathcal{G}_k} \lambda_{k,n} = 0,\\
&\lambda_{k,n} \left( t_k\left(\{\rho_{k,n}\}\right)- t_{\rm BA} \right)=0,~\forall (k,n),\\
& \sum\limits_{k=1}^K\sum\limits_{n\in \mathcal{G}_k} \rho_{k,n} - 1 =0, \lambda_{k,n}\geq 0,~t_k\left(\{\rho_{k,n}\}\right)\leq t_{\rm BA},\;\forall (k,n).
\end{aligned}
\right.
\end{equation}
In \eqref{eq:apKKTCP}, the second condition shows that $\exists (k,n),\; \lambda_{k,n}\neq 0$. Then, together with the first condition, 
it can be derived that $\mu\neq0$, and hence $\{\lambda_{k,n}\neq 0,\;\forall (k,n)\}$. In addition, according to the third condition in \eqref{eq:apKKTCP}, we have $\{t_{k,n}\left(\rho_{k,n}\right)- t_{\rm BA}=0,\; \forall (k,n)\}$.
In the next, by substituting $t_{k,n}\left(\rho_{k,n}\right)$ in \eqref{eq:apWL} into the above equation, we can derive \eqref{eq:OPBW}. 
Then, by substituting \eqref{eq:OPBW} into the condition $\sum\nolimits_{k=1}^K\sum\nolimits_{n\in \mathcal{G}_k} \rho_{k,n} - 1 =0$, \eqref{eq:BWS} can be derived. 
In \eqref{eq:BWS}, bisection search can be used to find $t_{\rm BA}^*$, as $B$ strictly decreases with $t_{\rm BA}$. Then, the optimal bandwidth-allocation scheme $\{\rho_{k,n}^*\}$ can be decided by \eqref{eq:OPBW}. 

\subsection{Proof of Lemma \ref{thm:OptimumConditions}}\label{apdx:thmOptimumConditions}
The KKT conditions of ({P5}) are used to show the sufficient and necessary conditions. First, the Lagrangian function is given by
\begin{equation}
\mathcal{L}\left(\{b_k\},t,\lambda,\mu\right) = t + \mu \left( \sum\limits_{k=1}^K b_k - N_{\rm p} \right) + \lambda \left(  \sum\limits_{k=1}^K\sum\limits_{n\in \mathcal{G}_k} \rho_{k,n}\left(b_k,t\right) - 1 \right),
\end{equation}
where  $\rho_{k,n}\left(b_k,t\right)$ is defined  in \eqref{eq:Rho}, and $\mu$ and $\lambda$ are multipliers. Then, KKT conditions are necessary to achieve the optimum, which are given by 
\begin{equation}\label{eq:KKTP5}
\left\{
\begin{aligned}
& \dfrac{ \partial \mathcal{L}\left(\{b_k\},t,\lambda,\mu\right) }{ \partial t } = 1+ \lambda \sum\limits_{k=1}^K \sum\limits_{n\in \mathcal{G}_k}\dfrac{ \partial \rho_{k,n}\left(b_k,t\right) }{ \partial t }  = 0, \\ 
&\dfrac{ \partial \mathcal{L}\left(\{b_k\},t,\lambda,\mu\right)}{ \partial b_k } = \mu + \lambda \sum\limits_{n\in \mathcal{G}_k}\dfrac{ \partial \rho_{k,n}\left(b_k,t\right) }{ \partial b_k }  = 0,\; \forall k, \\
&\lambda \left(  \sum\limits_{k=1}^K\sum\limits_{n\in \mathcal{G}_k} \rho_{k,n} - 1 \right) = 0,\sum\limits_{k=1}^K b_k = N_{\rm p}, \lambda \geq 0,\;\sum\limits_{k=1}^K\sum\limits_{n\in \mathcal{G}_k}\rho_{k,n} \leq 1. 
\end{aligned}
\right.
\end{equation}
From the first condition in \eqref{eq:KKTP5}, we have $\lambda \neq 0$. Then, the second conditions can be derived as 
\begin{equation}\label{eq:apC1}
 \sum\limits_{n\in \mathcal{G}_k}\dfrac{ \partial \rho_{k,n}\left(b_k,t\right) }{ \partial b_k }  = C,\; \forall k, 
\end{equation}
where $C = -\mu/\lambda$. Next, as $\lambda\neq 0$, the third condition in \eqref{eq:KKTP5} can be derived as
\begin{equation}\label{eq:apC2}
\sum\limits_{k=1}^K\sum\limits_{n\in \mathcal{G}_k} \rho_{k,n}\left(b_k,t\right) = 1.
\end{equation}
 As a result, the above two conditions in \eqref{eq:apC1} and \eqref{eq:apC2} together with the forth condition in \eqref{eq:KKTP5}, which are summarized in \eqref{eq:UniformGroupBWRates}, are necessary to achieve the minimal latency.
Furthermore, it is easy to show that only one solution is in \eqref{eq:UniformGroupBWRates}, which should be optimal. Hence, the conditions in \eqref{eq:UniformGroupBWRates} are sufficient and necessary to achieve the optimal latency.

\subsection{Proof of Lemma \ref{lma:Relation}}\label{apdx:lmaRelation}
First, two useful facts are listed below.
\begin{itemize}
\item Fact1:  The bandwidth allocation ratios $\{\rho_{k,n}\left(b_{k},t\right),\;\forall (k,n)\}$ defined in \eqref{eq:Rho} are monotonously decreasing function of the overall one-round latency $t$.
\item Fact2: $\rho_{k,n}\left(b_{k},t\right)$ is monotonously increasing function of the parametric-block length $b_k$ for all workers.
\end{itemize}
Based on Fact2, to maximize the feasible model size, second condition in ({P6}) should be
\begin{equation} \label{eq:Relation1}
\sum\limits_{k=1}^K\sum\limits_{n\in \mathcal{G}_k} \rho_{k,n}\left(b_k,t\right) = 1.
\end{equation}

Then, for  $t = T_1$,  denote the optimal solution of ({P6}) as $\{b_{k,1}^*\}$. The corresponding maximal feasible model size and optimal bandwidth allocation ratios are $n_{\rm p}^*(T_1)=\sum\nolimits_{k=1}^K b_{k,1}^*$ and $\{\rho_{k,n}^*\left(b_{k,1}^*,T_1\right)\}$, respectively. 

Next, assume any $T_2>T_1$, from Fact1, we have $\rho_{k,n}\left(b_{k,1}^*,T_2\right) < \rho_{k,n}^*\left(b_{k,1}^*,T_1\right)$ for all workers, 
which further shows that
\begin{equation}\label{eq:apdxRelation3}
\sum\limits_{k=1}^K\sum\limits_{n\in \mathcal{G}_k} \rho_{k,n}\left(b_{k,1}^*,T_2\right)< \sum\limits_{k=1}^K\sum\limits_{n\in \mathcal{G}_k} \rho_{k,n}^*\left(b_{k,1}^*,T_1\right)= 1.
\end{equation}
Furthermore, let
\begin{equation}\label{eq:Relation4}
b_{k,2}=b_{k,1}^*,\;k=2,3,...K.
\end{equation}
For group $\mathcal{G}_1$, we have
\begin{equation}\label{eq:Relation5}
\begin{aligned}
\sum\limits_{n\in \mathcal{G}_1}\rho_{k,n}\left(b_{1,2}, T_2\right)&= 1- \sum\limits_{k=2}^K\sum\limits_{n\in \mathcal{G}_k} \rho_{k,n}\left(b_{k,2},T_2\right),\\
&>1- \sum\limits_{k=2}^K\sum\limits_{n\in \mathcal{G}_k} \rho_{k,n}^*\left(b_{k,1}^*,T_1\right),\\
&= \sum\limits_{n\in \mathcal{G}_1}\rho_{k,n}^*\left(b_{1,1}^*,T_1\right),
\end{aligned}
\end{equation}
where the two equalities are due to \eqref{eq:Relation1} and the inequality is due to \eqref{eq:apdxRelation3} and \eqref{eq:Relation4}. According to Fact2 and \eqref{eq:Relation5}, we can show that $b_{1,2}>b_{1,1}^*$. That says $\sum\nolimits_{k=1}^K b_{k,2}>\sum\nolimits_{k=1}^K b_{k,1}^* = n_{\rm p}^*(T_1)$.
Furthermore, the maximal feasible model size for $t=T_2$ satisfies $n_{\rm p}^*(T_2)\geq \sum\nolimits_{k=1}^K b_{k,2}$. That says 
\begin{equation}
n_{\rm p}^*(T_2) > n_{\rm p}^*(T_1).
\end{equation}

\subsection{Proof of Lemma \ref{lma:ConvexityP6}}\label{apdx:lmaConvexityP6}
First, the objective of ({P6}) is convex and the the first constraint is a convex set. 
Then, given $t=T$,  from \eqref{eq:Rho}, all bandwidth-allocation ratios in the second condition can be written as 
\begin{equation}\label{eq:RhoT}
\rho_{k,n}\left(b_k\right) = \dfrac{b_k A_{\rm g}}{  \left(T - T_{\rm s} - T_{\rm ph} - \hat{t}_{k,n}\left(b_k\right)\right)R_{{\rm u},k,n} },\; \forall (k,n),
\end{equation} 
where $\hat{t}_{k,n}\left(b_k\right) = \dfrac{b_k D_{k,n} O}{f_{k,n}^{\rm c}}$.  In \eqref{eq:RhoT}, $\{\rho_{k,n}\left(b_k\right),\;\forall (k,n)\}$ can be linearly transformed from the convex function $f(x) = \dfrac{x}{1-ax}$ with $ax<1$ and $a>0$. Hence, $\{\rho_{k,n}\left(b_k\right),\;\forall (k,n)\}$ are convex, as linear transformation preserves convexity. Thereby, the second condition of ({P6}) is a convex set.

\bibliographystyle{ieeetr}
\bibliography{reference}

\begin{thebibliography}{10}

\bibitem{gesbert2019guest}
D.~Gesbert, D.~G{\"u}nd{\"u}z, P.~de~Kerret, C.~R. Murthy, M.~van~der Schaar,
  and N.~D. Sidiropoulos, ``Guest editorial special issue on machine learning
  in wireless communication-{P}art {I},'' {\em IEEE Journal on Selected Areas
  in Communications}, vol.~37, no.~10, pp.~2181--2183, 2019.

\bibitem{zhu2020toward}
G.~Zhu, D.~Liu, Y.~Du, C.~You, J.~Zhang, and K.~Huang, ``Toward an intelligent
  edge: Wireless communication meets machine learning,'' {\em IEEE Commun.
  Magazine}, vol.~58, pp.~19--25, Jan. 2020.

\bibitem{wang2018edge}
S.~{Wang}, T.~{Tuor}, T.~{Salonidis}, K.~K. {Leung}, C.~{Makaya}, T.~{He}, and
  K.~{Chan}, ``When edge meets learning: Adaptive control for
  resource-constrained distributed machine learning,'' in {\em Proc. IEEE Int.
  Conf. Comput. Comnun. (INFOCOM)}, (Honululu, USA), April 2018.

\bibitem{li2013parameter}
M.~Li, L.~Zhou, Z.~Yang, A.~Li, F.~Xia, D.~G. Andersen, and A.~Smola,
  ``Parameter server for distributed machine learning,'' in {\em Proc. NIPS
  Workshop on Big Learning}, (Lake Tahoe, USA), Dec. 2013.

\bibitem{li2014scaling}
M.~Li, D.~G. Andersen, J.~W. Park, A.~J. Smola, A.~Ahmed, V.~Josifovski,
  J.~Long, E.~J. Shekita, and B.-Y. Su, ``Scaling distributed machine learning
  with the parameter server,'' in {\em USENIX Symposium on Operating Systems
  Design and Implementation (OSDI)}, (Broomfield, USA), Oct. 2014.

\bibitem{wright2015coordinate}
S.~J. Wright, ``Coordinate descent algorithms,'' {\em Mathematical
  Programming}, vol.~151, no.~1, pp.~3--34, 2015.

\bibitem{li2014communication}
M.~Li, D.~G. Andersen, A.~J. Smola, and K.~Yu, ``Communication efficient
  distributed machine learning with the parameter server,'' in {\em Conference
  on Neural Information Processing Systems (NIPS)}, (Montr{\'e}al, Canada),
  Dec. 2014.

\bibitem{ho2013more}
Q.~Ho, J.~Cipar, H.~Cui, S.~Lee, J.~K. Kim, P.~B. Gibbons, G.~A. Gibson,
  G.~Ganger, and E.~P. Xing, ``More effective distributed ml via a stale
  synchronous parallel parameter server,'' in {\em Conference on Neural
  Information Processing Systems (NIPS)}, (Lake Tahoe, USA), Dec. 2013.

\bibitem{carreira2014distributed}
M.~Carreira-Perpinan and W.~Wang, ``Distributed optimization of deeply nested
  systems,'' in {\em Proc. Int. Workshop on Artificial Intelligence and
  Statistics (AISTATS)}, (Reykjavik, Iceland), April 2014.

\bibitem{zhang2016efficient}
Z.~Zhang, Y.~Chen, and V.~Saligrama, ``Efficient training of very deep neural
  networks for supervised hashing,'' in {\em IEEE Conference on Computer Vision
  and Pattern Recognition (CVPR)}, (Las Vegas, USA), June 2016.

\bibitem{lim2019federated}
W.~Y.~B. Lim, N.~C. Luong, D.~T. Hoang, Y.~Jiao, Y.-C. Liang, Q.~Yang,
  D.~Niyato, and C.~Miao, ``Federated learning in mobile edge networks: A
  comprehensive survey,'' {\em [Online]. Available:
  https://arxiv.org/pdf/1909.11875.pdf}, 2019.

\bibitem{chen2016revisiting}
J.~Chen, X.~Pan, R.~Monga, S.~Bengio, and R.~Jozefowicz, ``Revisiting
  distributed synchronous {SGD},'' {\em [Online]. Available:
  https://arxiv.org/abs/1604.00981.pdf}, 2016.

\bibitem{kamp2018efficient}
M.~Kamp, L.~Adilova, J.~Sicking, F.~H{\"u}ger, P.~Schlicht, T.~Wirtz, and
  S.~Wrobel, ``Efficient decentralized deep learning by dynamic model
  averaging,'' in {\em Joint European Conference on Machine Learning and
  Knowledge Discovery in Databases (ECML PKDD)}, (Dublin, Ireland), Sep. 2018.

\bibitem{chen2018lag}
T.~Chen, G.~Giannakis, T.~Sun, and W.~Yin, ``Lag: Lazily aggregated gradient
  for communication-efficient distributed learning,'' in {\em Conference on
  Neural Information Processing Systems (NIPS)}, (Montr{\'e}al, Canada), Dec.
  2018.

\bibitem{aji2017sparse}
A.~F. Aji and K.~Heafield, ``Sparse communication for distributed gradient
  descent,'' in {\em Conference on Empirical Methods in Natural Language
  Processing (EMNLP)}, (Copenhagen, Denmark), Dec. 2017.

\bibitem{zhu2019broadband}
G.~Zhu, Y.~Wang, and K.~Huang, ``Broadband analog aggregation for low-latency
  federated edge learning,'' {\em IEEE Trans. Wireless Commun.}, vol.~19,
  pp.~491--506, Oct. 2019.

\bibitem{amiri2019machine}
M.~M. Amiri and D.~Gunduz, ``Machine learning at the wireless edge: Distributed
  stochastic gradient descent over-the-air,'' {\em [Online]. Available:
  https://arxiv.org/abs/1901.00844, 2019.}, 2019.

\bibitem{ShiyuanmingAirComp}
K.~{Yang}, T.~{Jiang}, Y.~{Shi}, and Z.~{Ding}, ``Federated learning via
  over-the-air computation,'' {\em IEEE Trans. Wireless Commun.}, pp.~1--1,
  2020.

\bibitem{chen2019joint}
M.~Chen, Z.~Yang, W.~Saad, C.~Yin, H.~V. Poor, and S.~Cui, ``A joint learning
  and communications framework for federated learning over wireless networks,''
  {\em [online]. Available: https://arxiv.org/pdf/1909.07972.pdf}, 2019.

\bibitem{yang2019scheduling}
H.~H. Yang, Z.~Liu, T.~Q. Quek, and H.~V. Poor, ``Scheduling policies for
  federated learning in wireless networks,'' {\em IEEE Trans. Commun.},
  vol.~68, pp.~317--333, Sep. 2019.

\bibitem{zeng2019energy}
Q.~Zeng, Y.~Du, K.~K. Leung, and K.~Huang, ``Energy-efficient radio resource
  allocation for federated edge learning,'' {\em [online]. Available:
  https://arxiv.org/pdf/1907.06040.pdf}, 2019.

\bibitem{shi2019device}
W.~Shi, S.~Zhou, and Z.~Niu, ``Device scheduling with fast convergence for
  wireless federated learning,'' {\em [online]. Available:
  https://arxiv.org/pdf/1911.00856.pdf}, 2019.

\bibitem{ren2019accelerating}
J.~Ren, G.~Yu, and G.~Ding, ``Accelerating {DNN} training in wireless federated
  edge learning system,'' {\em [online]. Available:
  https://arxiv.org/pdf/1905.09712.pdf}, 2019.

\bibitem{yang2019energy}
Z.~Yang, M.~Chen, W.~Saad, C.~S. Hong, and M.~Shikh-Bahaei, ``Energy efficient
  federated learning over wireless communication networks,'' {\em [online].
  Available: https://arxiv.org/pdf/1911.02417.pdf}, 2019.

\bibitem{abad2019hierarchical}
M.~S.~H. Abad, E.~Ozfatura, D.~Gunduz, and O.~Ercetin, ``Hierarchical federated
  learning across heterogeneous cellular networks,'' {\em [online]. Available:
  https://arxiv.org/pdf/1909.02362}, 2019.

\bibitem{ruder2016overview}
S.~Ruder, ``An overview of gradient descent optimization algorithms,'' {\em
  [Online]. Available: https://arxiv.org/abs/1609.04747.pdf}, 2016.

\bibitem{parikh2014proximal}
N.~Parikh, S.~Boyd, {\em et~al.}, ``Proximal algorithms,'' {\em Foundations and
  Trends{\textregistered} in Optimization}, vol.~1, no.~3, pp.~127--239, 2014.

\bibitem{lang1995newsweeder}
K.~Lang, ``Newsweeder: Learning to filter netnews,'' in {\em Machine Learning
  Proceedings 1995}, pp.~331--339, Elsevier, 1995.

\end{thebibliography}

\appendices

\end{document}